\documentclass[preprint,floatfix,aps,prd,showpacs,nofootinbib,amsmath,amssymb,amsfonts,superscriptaddress]{revtex4-1}
\usepackage{graphicx}
\usepackage{color}
\usepackage{hyperref}
\usepackage{mathtools}
\usepackage{mathrsfs}
\usepackage{calrsfs}
\usepackage{comment}
\usepackage{lipsum}
\usepackage{footnote}
\usepackage[utf8]{inputenc}
\usepackage{multirow} 
\usepackage{subfigure}
\usepackage{varioref}
\usepackage[active]{srcltx}

\expandafter\ifx\csname package@font\endcsname\relax\else
\expandafter\expandafter
\expandafter\usepackage
\expandafter\expandafter
\expandafter{\csname package@font\endcsname}%
\fi
\hypersetup{
	colorlinks=true,       
	linkcolor=blue,          
	citecolor=blue,        
}
\newcommand*{\blue}{\textcolor{blue}}
\usepackage[section]{placeins}

\usepackage{fancybox}
\labelformat{figure}{Fig.~#1}

\newcommand{\rH}{r_{\rm H}}

\begin{document}
\title{Non-minimally coupled electromagnetic fields and observable implications for primordial black holes} 
\author{Susmita Jana} 
\email{susmitajana@iitb.ac.in}
\affiliation{Department of Physics, Indian Institute of Technology Bombay, Mumbai 400076, India}
\author{S. Shankaranarayanan}
\email{shanki@iitb.ac.in}
\affiliation{Department of Physics, Indian Institute of Technology Bombay, Mumbai 400076, India}
\begin{abstract}
General relativity (GR) postulates have been verified with high precision, yet our understanding of how gravity interacts with matter fields remains incomplete. Various modifications to GR have been proposed in both classical and quantum realms to address these interactions within the strong gravity regime. One such approach is non-minimal coupling (NMC), where the space-time curvature (scalar and tensor) interacts with matter fields, resulting in matter fields not following the geodesics. To probe the astrophysical implications of NMC, in this work, we investigate non-minimally coupled electromagnetic (EM) fields in the presence of black holes. Specifically, we show that primordial black holes (PBHs) provide a possible tool to constrain the NMC parameter. PBHs represent an intriguing cosmological black hole class that does not conform to the no-hair theorem. We model the PBH as a Sultana--Dyer black hole and compare it with Schwarzschild. We examine observables such as the radius of the photon sphere, critical impact parameter, and total deflection angles for non-minimally coupled photons for Schwarzschild and Sultana--Dyer black holes. Both the black hole space-times lead to similar constraints on the NMC parameter. For a PBH of mass $M=10^{-5} M_{\odot}$, the photon sphere will not be formed for one mode. Hence, the photons forming the photon sphere will be highly polarized, potentially leading to observable implications.


\end{abstract}
%
	%
	\maketitle
\section{Introduction}
\label{sec:Introduction}

General relativity (GR) describes the interaction of matter fields and gravity and manifests as a background geometry, quantified by the metric tensor \( g_{\mu\nu} \) and the covariant derivative \( \nabla_{\mu} \) associated with curved space-time. This influence, often termed minimal coupling between matter fields and gravity, can be nullified by constructing a freely falling frame. In such a frame, the laws of physics in GR mirror those described in the special theory of relativity, constituting the key postulate of GR known as Einstein's equivalence principle (EEP)~\cite{1977-Ni-PRL,1984-Gonner-FP}. The EEP encompasses three sub-principles: the weak EEP (or the universality of free fall), local Lorentz invariance, and local position invariance. Observable signatures of the EEP have undergone rigorous testing through various astrophysical phenomena, including the perihelion motion of Mercury, lunar-laser ranging, and more, with high precision~\cite{2014-Will-LRR}. Recent discoveries, such as the detection of gravitational waves by LIGO-VIRGO-KAGRA~\cite{2016-Abott.etal-PRL} and the imaging of supermassive black holes at the centers of galaxies~\cite{2019-Akiyama-Astrophys.J.Lett.}, have further strengthened the fundamental principles of GR.

Indeed, while astrophysical observations and experiments have provided robust validation of these postulates, their applicability is typically confined to the weak gravity limit~\cite{2007-Damour-Proc,2012-Speake.Will-CQG,2022-Shanki.Johnson-GRG}. However, the need for dark matter and dark energy to explain the current Universe, the elusive quantum theory of gravity, and its unification with the standard model of particle physics indicate the potential need for modifications to GR, particularly in the strong gravity regime~\cite{2014-Will-LRR}. The semiclassical theory of gravity has revealed that the quantum corrections of matter fields in curved space-time lead to effective actions featuring non-minimal coupling (NMC) to the geometry. More specifically, such couplings manifest as higher-order curvature invariants~\cite{1992-Vilkovisky-CQG,1994-Donoghue-PRD}. Recently, Ruhdorfer et al. did a systematic analysis and obtained an effective field theory of gravity coupled to the standard model of particle physics~\cite{2019-Ruhdorfer.etal-JHEP}. Their investigation showed the existence of non-minimal coupling of the standard model bosons with Riemann/Weyl curvature tensors leading to operators with mass dimension $6$. Intriguingly, the terms derived by these authors resemble the non-minimal coupling terms proposed by Prasanna~\cite{PRASANNA1971331}.

While classical and quantum theories of electrodynamics (ED) have demonstrated remarkable success, they are tested in the weak-gravity regime. Recent laboratory experiments, such as those involving X-rays, have tested quantum electrodynamics (QED)~\cite{2023-RIKEN-PRL}, yet these do not include gravity effects. Over the past four decades, numerous terrestrial and astrophysical tests have been devised to examine corrections to ED in the presence of gravitational fields~\cite{Drummond:1979pp,Daniels:1993yi,Daniels:1995yw,Latorre:1994cv, Dereli:2011mk,Horndeski:1976gi, Buchdahl:1979wi, Shore:2002gw, Balakin:2005fu, Linnemann:2021taq, Bergliaffa:2020ivp, Giani:2021jpt, Kushwaha:2020nfa,Kushwaha:2021csq}.
For instance, the initial study on the effects of non-minimal coupling focused on a Schwarzschild background~\cite{Drummond:1979pp}. Subsequent studies extended these investigations to include scenarios involving a static black hole with non-zero electric charge described by the Reissner--Nordström metric and a neutral rotating black hole described by the Kerr metric~\cite{Daniels:1993yi,Daniels:1995yw}. See also Refs.~\cite{Latorre:1994cv, Dereli:2011mk}. Various models of non-minimal coupling have been proposed by different authors~\cite{Horndeski:1976gi, Buchdahl:1979wi, Shore:2002gw, Balakin:2005fu, Linnemann:2021taq,Bergliaffa:2020ivp,Giani:2021jpt}. Recent studies have highlighted the potential of such coupling terms to address two unresolved issues in early Universe cosmology: the origin of primordial magnetic fields and the mechanism of baryogenesis~\cite{Kushwaha:2020nfa,Kushwaha:2021csq}. These findings suggest that the non-minimal coupling of Maxwell's theory with gravity may offer a promising avenue for resolving longstanding challenges in particle physics, gravity, and cosmology.

This leads us to the question: Will it be possible to constrain the NMC of EM fields? If yes, can 
such constraint be better than the current lower bounds from various astrophysical observations~\cite{Bedran1986AnEO,1994-LafranceRobertC-PRD,2014-Cheung-JHEP,2021-Bellazzini-JHEP,2020-deRham-PRD,2020-AccettulliHuber-PRD,2014-Camanho-JHEP}? In this work, we show that the primordial black holes (PBHs) provide a possible tool to constrain the NMC parameter~\cite{2021-Carr-Kuhnel-ARNPS,2021-Carr.etal-RPP}.
Specifically, we show that a PBH of mass around $10^{-5} M_{\odot}$ can provide a stringent constraint on the NMC parameter~$\lambda$. We compare the constraints on the NMC parameter from 
the next-generation Event Horizon Telescope (EHT).
Using these constraints, we evaluate the deflection angle of an NMC photon in the vicinity of a dynamical, spherically symmetric black hole described by the Sultana--Dyer metric~\cite{Sultana:2005tp} and a static black hole described by Schwarzschild space-time. To our knowledge, the detailed evaluation of the deflection angle in a black hole space-time to constrain the non-minimal coupling parameter is new. In Ref.~\cite{Prasanna:2003ix}, the authors used photon dispersion relation and computed the photon's arrival time in Schwarzschild space-time. Considering signals from radar ranging past the Sun, the authors found $\lambda \sim 1.1 \times 10^{20}~\rm{cm}^2$, which is about three orders of magnitude more stringent than the one obtained in Ref.~\cite{Bedran1986AnEO}. We show that deflection angle analysis improves the bound on $\lambda$ by 10 \emph{orders of magnitude} for a stellar mass black hole and is consistent with the constraint obtained from a binary pulsar. 

Cosmological dark matter is thought to be a complex mixture of currently hypothetical entities such as primordial black holes (PBHs) and predominantly unknown weakly interacting massive particles (WIMPs) like neutrinos and axions~\cite{Misiaszek:2023sxe}. In the past few years, after LIGO detected black holes in previously unexplored mass ranges, there has been a renewed interest in PBHs~\cite{2021-Carr-Kuhnel-ARNPS,2021-Menaetal-Frontiers}. The PBH was introduced by Hawking half a century ago~\cite{1971-Hawking-MNRAS}, with subsequent recognition that these BHs could constitute the majority of matter in the Universe~\cite{Carr:1974nx,1975-Chapline-Nature,1979-Zeldovich_Novikov-NASA}. PBHs are not subject to the constraints of big bang nucleosynthesis, so they are effectively non-baryonic DM~\cite{2021-Carr-Kuhnel-ARNPS}; while conclusive evidence remains elusive, diverse and accurate observations have revealed at least four mass windows potentially accommodating primordial black holes: the $10^{-16} - 10^{-10} M_{\odot}$ range and the $10^{-6} \blue{-} 10^{-5} M_{\odot}$ range for small black holes, the $10 \blue{-} 10^{3} M_{\odot}$ range for medium-sized black holes, and the hyper-massive range beyond~\cite{2021-Carr-Kuhnel-ARNPS}. It has recently been suggested that the outer Solar System harbors a PBH of mass $\sim 10^{-6} \, M_{\odot}$~\cite{Scholtz:2019csj}.

A curious reader will then wonder why PBHs can provide stringent constraints for the NMC parameter compared to astrophysical BHs of around $10 - 10^2 \, M_{\odot}$. For a gravitating object of mass $M$ and size (or radius) $L$, two parameters $(\epsilon, {\cal K})$ serve as useful indicators to categorize gravitational fields~\cite{2021-GWIC3G-Arxiv,2022-Shanki.Johnson-GRG}. $\epsilon$ is a dimensionless parameter, while ${\cal K}$ is the square root of the Kretschmann scalar:
\[
\epsilon \sim GM/(L c^2) \, , \, {\cal K} = \left( R_{\mu \nu\sigma \rho}~R^{\mu \nu\sigma \rho} \right)^{1/2} \, ,
\] 
where \( R_{abcd} \) is the Riemann curvature tensor. Thus, \( \epsilon \) indicates the gravitational strength, helping to distinguish between the weak and strong gravity regimes. For instance, the Solar System, characterized by very small \( \epsilon \) values, clearly operates within the weak gravity regime, while BHs, where \( \epsilon \) approaches or equals one near the event horizon, imply a strong gravity regime. {Using the definition of the tidal tensor 
$\Phi_{\mu\nu} \equiv R_{\mu 0 \nu 0}$, we see that the Kretschmann scalar $K$ 
measures the tidal intensity.} The Kretschmann scalar ($\cal{K}$) remains finite everywhere outside the event horizon of a BH. Specifically, ${\cal K} \propto L^{-2}$ at the event horizon of a Schwarzschild BH. For a solar mass BH, ${\cal K} \sim 10^{-6}~{\rm m}^{-2}$, while for a PBH of $10^{-3} M_{\odot}$, ${\cal K} \sim 1~{\rm m}^{-2}$. This implies that smaller BHs exhibit increasingly strong curvature near their horizons. Hence, NMC will have a significant impact on PBHs. In other words, PBHs can provide stringent constraints for the NMC parameter. 

PBHs, as potential dark matter candidates, represent an intriguing class of cosmological BHs that emerge from density fluctuations in the early Universe. One of the distinguishing characteristics of PBHs is that they may not be stationary or asymptotically flat \footnote{In principle, all astrophysical black holes are non-stationary. However, the timescales of various astrophysical processes are so large that these black holes can be effectively modeled as isolated, stationary black holes in general relativity. In contrast, for primordial black holes (PBH) formed during the early Universe, these assumptions are not valid~\cite{Carr:1974nx,1975-Chapline-Nature,1979-Zeldovich_Novikov-NASA}}. This differentiates them from the isolated BH solutions in general relativity, such as the Schwarzschild or Kerr black holes, defined under stationarity and asymptotic flatness. Instead, PBHs could be dynamic and evolving, interacting with the expanding Universe and other matter, influenced by accretion and possibly emitting Hawking radiation~\cite{2021-Carr-Kuhnel-ARNPS,2022-Xavier.etal-PRD}.

This implies that PBHs do not necessarily conform to the no-hair theorem, which states that black holes in equilibrium states are fully described by just three externally observable parameters: mass, electric charge, and angular momentum. The theorem presumes a stationary and isolated black hole, conditions not typically applicable to cosmological black holes like PBHs. In cosmological contexts, black holes could have additional characteristics derived from their interactions and evolution in a non-flat space-time~\cite{2024-Sunny.etal-Arxiv}.

The possible deviation from the no-hair theorem in cosmological settings like those involving PBHs suggests complexities in their properties and behaviors, reflecting interactions with the cosmic environment and potentially making them distinct in terms of observational signatures from black holes that are purely described by general relativity. Sultana and 
Dyer (SD) obtained an exact spherically symmetric black hole solution in expanding cosmological space-time~\cite{Sultana:2005tp}. An SD black hole is asymptotically Friedmann--Lemaitre--Robertson--Walker~(FLRW)~\cite{2014-Faraoni-PRD,2018-Faraoni-Universe}. Thus, the SD black hole can mimic a PBH in cosmological scenarios~\cite{2018-Faraoni-Universe}. 

As mentioned earlier, we study the propagation of photons that are non-minimally coupled to gravity. We show that the two dispersion relations corresponding to the two polarizations differ from Schwarzschild and Kerr (photon confined in equatorial plane) [cf. 
Equation~\eqref{eq:dispersion_relation_sch}]. In the case of Schwarzschild and Kerr, the two dispersions are quadratic, and the modifications to the dispersion relation occur in the $p_{(3)}$ component of the momentum. However, in the case of Sultana--Dyer, the modification to the dispersion relation [cf. Equations~\eqref{eq:SD-DRelation-NLocal1}, \eqref{eq:SD-DRelation-NLocal2-part1}, and \eqref{eq:SD-DRelation-NLocal2-part2}] occurs in all the components of the momentum. Further, one of the dispersion relations is quadratic, and the other is quartic. Interestingly, the quartic dispersion relation reduces to two quadratic dispersion relations for small values of $\tilde{\lambda}:= \lambda/\rH^2$. We show that the radius of photon sphere $(r_{0})$ and the corresponding value of impact parameter~$(b_0)$ are different in SD and Schwarzschild space-times 
for a fixed value of $\tilde{\lambda}$. Even these values differ for different dispersion relations in the same space-time. 

Then, from the dispersion relations, we derive the expressions for deflection angle $(d\phi/dr)$ and evaluate total deflection angle~$(\Delta \phi)$ for every mode of polarization in a particular space-time (SD or Schwarzschild) and compare the results. We show that $\Delta \phi $ behaves differently as a function of $r$, the radial distance from the black hole in the SD and Schwarzschild cases. We compute the total deflection angle for non-minimally coupled photons near Schwarzschild and Sultana--Dyer black holes and show that both these black holes have similar signatures with regard to the total deflection angle as well as the radius of the photon sphere. Interestingly, for a PBH of mass $M=10^{-5} M_{\odot}$, the photon sphere will not be formed for one mode. Thus, the total deflection angle $\Delta \phi$ of NMC photons near Schwarzschild and SD space-times will have a contribution \emph{only} from the other modes. In other words, the photons forming the photon sphere will be highly polarized, potentially leading to observable implications.

This paper is organized as follows: In Section~\ref{sec:Model}, we discuss the non-minimal coupling model of the electromagnetic fields with gravity and briefly discuss the procedure to obtain the photon dispersion relation in arbitrary 4D space-time. In Section~\ref{sec:Sultana-Dyer metric}, we briefly discuss the properties of the Sultana--Dyer black hole and the conserved quantities in this space-time. In Section \ref{sec:Dispersionrelations}, we obtain the modified dispersion relations in the local inertial frame for the Schwarzschild and Sultana--Dyer black holes. In Section \ref{sec:Def-Angle}, we evaluate the expressions for the deflection angles and total deflection angles for all polarization modes as a function of distance from the Schwarzschild (two polarization modes) and Sultana--Dyer (three polarization modes) black holes. We also obtain the constraint on the non-minimal coupling parameter $\lambda$. In Section \ref{sec:Comparison}, we obtain the quantitative difference in the total deflection angle for the Sultana--Dyer black hole from the Schwarzschild. In Section~\ref{sec:conclusion}, we conclude by discussing the implications of our work for future observations. Appendixes \ref{sec:Dispersion relation with Kerr metric}--\ref{sec:lambda from EM tensor} contain calculation details.

In this work, we use the 
$(-, +, +, +)$ signature for the 4D space-time metric. Greek alphabets denote the four-dimensional space-time coordinates, and Latin alphabets denote the three-dimensional spatial coordinates.

\section{The model}
\label{sec:Model}

The most general action of non-minimal coupling of electromagnetic (EM) fields~\cite{Balakin:2005fu}
is 
\begin{align}
\label{eq:S_Balakin}
\mathcal{S}_{NC} = \int d^{4}x \sqrt{-g} \left[ \frac{R}{\kappa} \ - \frac{1}{4} \ F_{\mu\nu}F^{\mu\nu} \  + \  \chi ^{\mu \nu \alpha \beta} \ F_{\mu\nu}F_{\alpha\beta} \right]
\end{align}
where,
\begin{align}
\label{eq:chi}
\chi ^{\mu \nu \alpha \beta} = \ \frac{q_{1}R}{2} \left(g^{\mu \alpha} g^{\nu \beta} - g^{\mu \beta} g^{\nu \alpha}  \right) \ + \ \frac{q_{2}}{2} \left(R^{\mu \alpha} g^{\nu \beta} - R^{\mu \beta} g^{\nu \alpha} \ + \ R^{\nu \beta} g^{\mu \alpha} - R^{\nu \alpha} g^{\mu \beta}\right)
  + \frac{q_{3}}{2} R^{\mu\nu\alpha\beta} \, .
\end{align}
and consists of coupling between the EM field tensor, the Riemann and Ricci tensors, and the Ricci scalar. $q_1, q_2$ and $q_3$ are the coupling parameters.
The above action is invariant under CPT and general coordinate transformations; however, it violates Einstein's equivalence principle~\cite{Prasanna:2003ix}. In our work we assume $q_{1} \simeq q_{2} \simeq q_{3} \equiv \lambda$.

In Ref.~\cite{Drummond:1979pp},  Drummond and Hathrell used Schwinger-de Witt's effective action approach to obtain the non-minimal coupling terms. To obtain the effective action, one separates the degrees of freedom into two parts --- heavy($h$) and light($l$) --- corresponding to their field modes in the momentum space. The 
influence of heavy fields on light-particle scattering at low energies leads to the following effective Lagrangian:
\begin{equation}
\label{leffdef}
\exp\left[i \int d^4x \; {\cal L}_{\rm eff}(\ell) \right] 
= \int {{\cal D}}h \; \exp\left[ i \int d^4x \;
{\cal L}(\ell,h) \right] \, ,
\end{equation}
{where ${\cal D}$ is the measure in the field space and integration is over the heavy field}. It is important to note that this is a low-energy effective action {obtained} by integrating out the heavy fields. In the case of Drummond and Hathrell, this corresponds to photon energies small compared to the electron mass, i.e., $\hbar \omega \ll m_e c^2$. In this limit, they found that the non-minimal coupling parameters are of the same order, i.e., $q_{1} \simeq q_{2} \simeq q_{3} \equiv \lambda$~\cite{Drummond:1979pp}. It is important to note that Drummond and Hathrell's analysis is suitable for electromagnetic waves with energies much less than $0.5~{\rm MeV}$, and their analysis is not trustable for high-energy photons (hard x-rays).

%
%

Varying the action (\ref{eq:S_Balakin}) with respect to the gauge field $A_{\mu}$ and taking the three coupling parameters to be the same ($q_1 = q_2 = q_3 = \lambda$), we get:
\begin{align}
\nonumber
(1 - 4\, \lambda R) \, \nabla_{\nu}F^{\mu\nu} = & 6 \lambda  (\nabla_{\nu} R)F^{\mu\nu} - 8 \lambda (\nabla_{\alpha} R^{\mu}\, _{\beta})F^{\alpha\beta}  \\
\label{eq:EoM_withRicci}
& + 4\lambda\left[\,  R^{\mu}\, _{\alpha}(\nabla_{\nu}F^{\alpha\nu}) + R^{\nu}\, _{\alpha}(\nabla_{\nu}F^{\alpha\mu})\, \right] + 2\lambda R^{\mu\nu}\, _{\alpha\beta}(\nabla_{\nu}F^{\alpha\beta}) \, .
\end{align}
%
%
The RHS of the above equation is a coordinate-dependent 
quantity and lead to a modified dispersion relation. In the case of Maxwell's theory, we will have the same dispersion relation ($p^{\mu}p_{\mu} = 0$) for both polarization modes.
The first two terms in the RHS act as the mass term, while the last two terms  in the RHS modify the kinetic term. As the coupling terms are linearly related to EM fields, the non-minimal coupling leads to 
mixing of the two polarization modes. 

For isolated, static black holes, like Schwarzschild (de Sitter) or Kerr (de Sitter) metric, that satisfy $R_{\mu\nu} \propto g_{\mu\nu}$, only the last term in the RHS of the above expression is non-zero. However, for black holes in an expanding universe, like the Sultana-Dyer metric, $R_{\mu\nu} \neq \Lambda g_{\mu\nu}$. Hence, we retain all terms in the RHS of the above expression. As mentioned in the Introduction, since ${\cal K} \propto L^{-2}$ at the event horizon of a Schwarzschild BH, the NMC term will have significant effect on PBH.

\subsection{Photon dispersion relation}
\label{subsec:Photon_dispersion_relation}

To obtain the photon trajectory, it is advantageous to use the tetrad basis of four-linearly independent vector fields ($e^{\mu}\, _{(\nu)}$ where $\mu = 0, 1, 2, 3$). The enclosure in the parenthesis distinguishes tetrad indices from the tensor indices. The tetrads $e^{\mu}\, _{(\nu)}$ or its inverse $e_{\mu}\, ^{(\nu)}$ satisfy the relation, $e^{\mu}\, _{(\alpha)}e^{\nu}\, _{(\beta)} \, g_{\mu\nu} = \tilde{\eta}_{(\alpha)(\beta)}$. As in the literature, we assume that  $e^{\mu}\, _{(\nu)}$ are orthonormal and hence, $\tilde{\eta}_{(\alpha)(\beta)}$ represents metric corresponding to flat space-time with the diagonal elements $\left(\,-1, +1, +1, +1 \, \right)$ and $g_{\mu\nu}$ denotes the metric tensor describing an arbitrary curved spacetime~\cite{1973-MTW-Book}. The tetrads and their inverse are related as: $e^{\mu}\, _{(\alpha)}e_{\nu}\, ^{(\alpha)} = \delta^{\mu}_{\nu}$, where $\delta^{\mu}_{\nu}$ is the Kronecker delta function. If  $p_{(\mu)}$ corresponds to the photon momentum in the local inertial frame, this is related to momentum in the non-inertial frame, $p_{\nu}$ by the relation:  $p_{(\nu)}  =  e^{\mu}\, _{(\nu)} \ p_{\mu}$.

In the case of time-dependent black holes, we consider the variations of the curvature (${\cal K})$ along the radial coordinate  and the time (${T}$) is long compared to the EM waves reduced wavelength and period, i.e.
\[
{\cal K} \gg  \lambdabar = 1/k , {T} \gg 1/\omega  \, .
\]
Under the Eikonal approximation, we can locally treat the EM waves as planar and monochromatic~\cite{Prasanna:2003ix}:
\begin{align}
\label{eq:eikonal_approximation}
\nabla_{(\mu)}F^{(\mu)(\nu)}  = p_{(\mu)} \, F^{(\mu)(\nu)} \, .
\end{align}
Thus, in the local inertial frame, the Bianchi identity of the EM field tensor and Eq.~(\ref{eq:EoM_withRicci}) reduces to:
\begin{eqnarray}
\label{eq:Bianchi_identity in local}
& & p_{(\alpha)}F_{(\mu)(\nu)} + p_{(\mu)}F_{(\nu)(\alpha)} + p_{(\nu)}F_{(\alpha)(\mu)} = 0
\end{eqnarray}
and,
\begin{align}
\nonumber
& (1 - 4\, \lambda \, R) \, p_{(\nu)}F^{(\mu)(\nu)}\,  = 2 \lambda\,p_{(\nu)}F^{(\alpha)(\beta)}\, R^{(\mu)(\nu)}\, _{(\alpha)(\beta)} \,
- 8 \lambda \,F^{(\alpha)(\beta)} \, \nabla_{(\alpha)} R^{(\mu)}\, _{(\beta)}   \\
\label{eq:EoM_withRicci_eikonal}
& + 4 \lambda \left[\,p_{(\nu)}F^{(\alpha)(\nu)}\,  R^{(\mu)}\, _{(\alpha)}\, + \, p_{(\nu)}F^{(\alpha)(\mu)} \, R^{(\nu)}\, _{(\alpha)}\, \right] +
\, 6 \lambda \,F^{(\mu)(\nu)}  \, \nabla_{(\nu)} R \,
\end{align}
Note that the second 
and the last terms in RHS are independent of the 4-momentum, $p_{(\nu)}$. For high-momentum photons, the contribution of these terms can be neglected compared to the other two terms that are linear in order. Thus, the above expression reduces to:
{\footnotesize
\begin{align}
\label{eq:EoM_withRicci_eikonal_Simplified}
(1 - 4\, \lambda R) \, p_{(\nu)}F^{(\mu)(\nu)}\,  \simeq \,
2 \lambda R^{(\mu)(\nu)}\, _{(\alpha)(\beta)}(p_{(\nu)}F^{(\alpha)(\beta)}) +
 4 \lambda \left[\,  R^{(\mu)}\, _{(\alpha)}(p_{(\nu)}F^{(\alpha)(\nu)}) + R^{(\nu)}\, _{(\alpha)}(p_{(\nu)}F^{(\alpha)(\mu)})\, \right] 
\end{align}
}
Setting the space-time indices 
$(\mu,  \nu) $ to 3-space index $(j,k)$,  respectively and setting $\alpha = 0$ (the time index) --- the Bianchi identity in local inertial frame~\eqref{eq:Bianchi_identity in local} becomes:
\begin{eqnarray}
\label{eq:Bianchi_identity in local2}
p_{(0)}F_{(k)(j)} \, = \, p_{(k)}F_{(0)(j)} - p_{(j)}F_{(0)(k)}  
\end{eqnarray}
{In Eq.~\eqref{eq:EoM_withRicci_eikonal_Simplified} we split the dummy index into time ($0$),  spatial parts($k$) and fixing the space-time index $\mu$ to the 3-space index $j$ and combining Eq.~\eqref{eq:Bianchi_identity in local2} with them leads to the respective equations:  }
{\footnotesize
\begin{align}
\label{eq:EoM_withRicci_eikonal_Simplified_zero}
& \biggr[  (1 - 4\, \lambda R) \, p_{(k)}p^{(0)} -  4 \lambda \left(\,  
R^{(0)}\, _{(\alpha)}p_{(k)}p^{(\alpha)}
-R^{(0)}\, _{(k)}  p_{(\nu)}p^{(\nu)}  - R^{(\nu)}\, _{(k)}p_{(\nu)}p^{(0)}\, \right)  \\ 
& \qquad \qquad \qquad \qquad - 4 \lambda \, R^{(0)(\nu)(\alpha)}\, _{(k)}p_{(\nu)}p_{(\alpha)}   \biggr] F^{(k)(0)} = 0 \nonumber \\
\nonumber
& \biggr[\, (1 - 4\, \lambda R) \,\left(  p_{(\nu)}p^{(\nu)}\delta^{(j)}_{(k)} - p_{(k)}p^{(j)}  \right) + \, 4 \lambda \, \biggr( \,  
R^{(j)(\alpha)}\, p_{(k)}p_{(\alpha)} + R^{(\nu)(\alpha)}\, p_{(\nu)}p_{(\alpha)} \, \delta^{(j)}_{(k)} 
- R^{(j)}\, _{(k)} \,  p_{(\nu)}p^{(\nu)}   \\
\label{eq:EoM_withRicci_eikonal_Simplified_j}
& - R^{(\nu)}\, _{(k)}\, p_{(\nu)}p^{(j)} \, \biggr)  + 4 \lambda R^{(j)(\nu)(\alpha)}\, _{(k)}p_{(\nu)}p_{(\alpha)} \, \biggr]  F^{(k)(0)} = 0
\end{align}  
}
Combining Eqs.\eqref{eq:EoM_withRicci_eikonal_Simplified_zero} , \eqref{eq:EoM_withRicci_eikonal_Simplified_j} and \eqref{eq:Bianchi_identity in local2} leads to the following equation of motion:
{\footnotesize
\begin{align}
\nonumber
& \biggr[\, (1 - 4\, \lambda R)\,  p_{(\nu)}p^{(\nu)}\delta^{(j)}_{(k)}
+ 4 \lambda \biggr(R^{(j)(\alpha)}\, p_{(k)}p_{(\alpha)} + R^{(\nu)(\alpha)}\, p_{(\nu)}p_{(\alpha)} \, \delta^{(j)}_{(k)} 
- R^{(j)}\, _{(k)} \,  p_{(\nu)}p^{(\nu)} - R^{(\nu)}\, _{(k)}\, p_{(\nu)}p^{(j)} \\
\label{eq:Final-EoM-withRicci}
&  + \, \epsilon^{(j)}\, _{(k)}\biggr) + \frac{p^{(j)}}{p_{(0)}}\, 4 \lambda \, \biggr(\, R^{(0)}\, _{(\alpha)}p_{(k)}p^{(\alpha)}
-R^{(0)}\, _{(k)}  p_{(\nu)}p^{(\nu)}  - R^{(\nu)}\, _{(k)}p_{(\nu)}p^{(0)} + \epsilon^{(0)}\, _{(k)} \, \biggr)
\,\biggr] F^{(k)(0)}  = 0
\end{align}
}
where, 
\begin{align}
\label{eq:define1}
\epsilon^{(\alpha)} \, _{(\beta)} &\equiv R^{(\alpha)(\mu)(\nu)} \, _{(\beta)} \, p_{(\mu)}p_{(\nu)}.
\end{align} 
%
The above equation of motion is for an arbitrary space-time.  By choosing a coordinate system,  it is always possible to obtain the dynamical and the constraint equations. 
Interestingly, we see that the terms containing $p^{(j)}/p^{(0)}$ can potentially lead to higher-order dispersion relations.  As we will see,  this is the case for the Sultana-Dyer black hole.  Thus, for different space-times, the above equation of motion \eqref{eq:Final-EoM-withRicci}  will yield different dispersion relations and potentially provide constraints on $\lambda$ for that particular space-time.

The constraints on $\lambda$ are obtained in the literature by considering static black hole space-times, like Schwarzschild and {slowly rotating Kerr} \cite{Drummond:1979pp, Prasanna:2003ix, 2023-Johnson.Jana.Shanki-gr-qc}.  In this work, we obtain constraints on $\lambda$ by considering a spherically symmetric black hole in cosmological space-time, commonly referred to as Sultana-Dyer (SD) metric~\cite{Sultana:2005tp}. 
To our knowledge, such an analysis has not been done, and we show that this provides a stringent constraint on the non-minimal coupling parameter $\lambda$, {specifically for PBHs}. We also show the radius of the photon sphere, and corresponding critical impact parameter have similar values in both SD and Schwarzschild space-time for a fixed value of $\lambda$. We discuss the implications on these quatities for PBH also.
\section{Sultana-Dyer black hole}
\label{sec:Sultana-Dyer metric}

Sultana and Dyer obtained an exact spherically symmetric black hole solution in expanding cosmological space-time\cite{Sultana:2005tp}:
\begin{align}
\label{eq:metric_SDyer_non-diagonal1}
ds^{2} = \left(\frac{\eta}{\eta_{0}}\right)^{4}\left[ - \left(1-\frac{2GM}{r} \right)\ d\eta^{2} \ + \frac{4GM}{r}\ d\eta dr+ \left(1+\frac{2GM}{r} \right) dr^{2}\ + \ r^{2} \ d\Omega^{2} \right] \, 
\end{align}
 where $M$ is the mass of the non-rotating black hole and 
 $\eta_{0}$ is the conformal time at current epoch, $t_{0} \sim 1/H_0$, and $H_0$ is Hubble constant. The conformal time $\eta$ is related to the comoving or cosmic time $t$ via the relation  $dt =  d\eta \, a(\eta)$ where the conformal scale factor for the above metric is $ a(\eta) = \left(\eta/\eta_{0}\right)^2$. Integrating the relation connecting conformal time and cosmic time, we get $t=\eta^3/(3\, \eta_{0}^2)$, and $t_{0} = \eta_0/3$.
We list below some of the properties of the Sultana-Dyer space-time:
\begin{enumerate}
\item The above line element corresponds to a black hole of mass $M$ in a 
spatially flat FLRW universe with scale factor $a(t) \propto t^{2/3}$. This 
corresponds to black holes in the matter-dominated epoch. 
\item Sultana-Dyer is sourced by two non-interacting perfect fluids --- null dust and normal dust. The stress-energy tensor is 
$T_{\mu \nu}=T_{\mu \nu}^{(1)}+T_{\mu \nu}^{(2)}$, 
where $T_{\mu \nu}^{(1)}=\rho_0 u_{\mu} u_{\nu}$ describes the normal dust with density $\rho_{0}$ and $u^{\alpha} u_{\alpha}= -1$
and $T_{\mu \nu}^{(2)}=\rho_{n} k_{\mu} k_{\nu}$ describes a null dust with density $\rho_{n}$ and $k^{\alpha} k_{\alpha}=0$. 
\item 
The Kretschmann scalar for the line element~\eqref{eq:metric_SDyer_non-diagonal1} is:
\begin{align}
\label{eq:Kretschmann-SD-dimensionfull}
R^{\mu\nu\alpha\beta}R_{\mu\nu\alpha\beta}  \equiv \mathcal{K} = 48\,  \eta_{0}^8 \, \left( \frac{8 M^2}{{r^4 \eta^{10}}} + \frac{M^2}{r^6 \eta^{8}} - \frac{4 M   (4 M+r)}{r^3 \eta^{11}} + \frac{5 (2 M+r)^2}{r^2 \eta^{12}}\right) \, .
\end{align}
It implies curvature singularities occur at  $\eta = 0$ and $r = 0$. The singularity at $\eta = 0$ is spacelike for $r > 2\, GM \equiv \rH$, timelike for $r < \rH$ and null for $r = \rH$~\cite{2009-Faraoni-PRD}.  The singularity at $r = 0$ is spacelike and surrounded by the event horizon. 
\item The energy density of the dust is positive only in the region 
\begin{equation}
\label{eq:SD-EnerDen}
\eta  < \frac{r (r + \rH)}{\rH}
\end{equation}
At  $r= \rH$, the energy conditions for $\eta < 2 \rH$ are satisfied everywhere. In other words, after this time $\eta$, those particles closest to the event horizon become superluminal~\cite{Sultana:2005tp}. As we will see, this condition translates to a constraint on $\lambda$.

\item To understand the properties of the horizons, we rewrite the time coordinate $\eta$ as 
\begin{align}
\label{eq:eta-tbar}
\eta =\bar{t} + 2M \ln \left(\frac{r}{2\, GM} -1 \right)
\end{align}
in the line element~\eqref{eq:metric_SDyer_non-diagonal1}. This leads to:
\begin{align}
\label{eq:metric_SDyer_diagonal}
d\bar{s}^{2} = a(\bar{t},r)^{2}\left[ - \left(1-\frac{2GM}{r} \right)\ d\bar{t}^{2} \ + \left(1 - \frac{2GM}{r} \right)^{-1} dr^{2}\ + \ r^{2} \ d\Omega^{2} \right] \, 
\end{align}
From the above line element, it is evident that the Sultana-Dyer metric is a black hole obtained by the conformal transformation of the Schwarzschild black hole. As the conformal transformation preserves the causal structure, $r = 2GM \equiv \rH$ remains the event horizon. Interestingly, since the line-element is conformally invariant, there exists a conformal Killing vector $\zeta = \partial_{t}$ which is the Killing vector on Schwarzschild space-time and satisfies the following relation~\cite{2007-Saida-CQG},
\begin{align}
\mathcal{L}_{\zeta}\, g_{\mu\nu} = \left(\mathcal{L}_{\zeta}\, \ln a^{2}\right)\, g_{\mu\nu}
\end{align}
where, $\mathcal{L}_{\zeta}\, \ln a^{2} = 4/\eta$.
The \textit{apparent horizon} is given by~\cite{2007-Saida-CQG,2014-Faraoni-PRD}:
\begin{align}
\label{eq:apparent-horizon-1}
r_{1} & = - GM + \frac{-\, \eta + \sqrt{\eta^{2} + 24 \, GM\eta + 16 \, (GM)^{2}}}{4} \\
\label{eq:apparent-horizon-2}
r_{2} & = \frac{\eta}{2}
\end{align}
where, $r_{1} < r_{2}$. From the energy condition \eqref{eq:SD-EnerDen},  it is clear that $r_{1} < r_{2} < \rH$, implying that the \textit{apparent horizon} lies inside the event horizon and asymptotically tends to event horizon.

\item  Writing the line element \eqref{eq:metric_SDyer_non-diagonal1} in non-geometrized unit leads  to:
\begin{align}
\label{eq:metric_SDyer_non-diagonal-non-geometric}
ds^{2} = \left(\frac{\eta}{\eta_{0}}\right)^{4}\left[ -\left(1-\frac{\rH}{ r }\right)c^2\ d\eta^{2} \ + \frac{2  \rH}{r} c\,  d\eta dr+ \ \left(1+\frac{\rH}{ r }\right)\ dr^{2}\ + \ r^{2} \ d\Omega^{2} \right].
\end{align}

Substituting the following dimensionless variables:
\begin{align}\label{eq:dimension-less coordinate}
\tilde{r} =  \frac{r }{\rH} \, ; \quad
\tilde{\eta} \equiv c \, \eta/\rH \, ; \quad
\tilde{\eta}_{0} \equiv c \, \eta_{0}/\rH 
\end{align}
in the Sultana-Dyer metric \eqref{eq:metric_SDyer_non-diagonal-non-geometric}, we have
\begin{align}
\label{eq:metric_SDyer_non-diagonal4}
ds^{2}&= \rH^2\ \tilde{\eta}^4 \left[-\left(1-\frac{1}{\tilde{r}}\right)\   d\tilde{\eta}^2 + \frac{2}{\tilde{r}}  \  d\tilde{\eta}d\tilde{r} + \left(1+\frac{1}{\tilde{r}}\right)d\tilde{r}^2 + \tilde{r}^2 \ d\Omega^2\right]  = \rH^2 \  d\tilde{s}^{2} 
\end{align}
where, 
\begin{eqnarray}
\label{eq:dimensionless-SDyer-metric}
 d\tilde{s}^{2} &=& \tilde{\eta}^4 \left[-\left(1-\frac{1}{\tilde{r}}\right)\   d\tilde{\eta}^2 + \frac{2}{\tilde{r}}  \  d\tilde{\eta}d\tilde{r} + \left(1+\frac{1}{\tilde{r}}\right)d\tilde{r}^2 + \tilde{r}^2 \ d\Omega^2\right] 
\end{eqnarray}
%
\end{enumerate}

In the rest of the work, we will use the dimensionless line-element $d\tilde{s}$ 
defined in Eq.~\eqref{eq:dimensionless-SDyer-metric}. This choice is suitable because the affine parameter ($\tau$) and conserved quantities are dimensionless. 

\subsection{Conserved quantities from symmetries of the metric}

Although Sultana-Dyer metric \eqref{eq:dimensionless-SDyer-metric} is time-dependent, the space-time has conserved quantities. Besides the spherical symmetry, the metric is invariant under conformal transformations. Taking into 
account of the conformal symmetry; the following quantity is conserved 
w.r.t dimensionless affine parameter $\tau$~\cite{Sultana:2005tp,2009-Faraoni-PRD}:
\begin{align}
\label{eq:tilde-E-dimensionless}
p_{0} \equiv -\tilde{E}_{\rm SD} &= \tilde{\eta}^4\left(- \left(1-\frac{1}{\tilde{r}}\right) \, \frac{d\tilde{\eta}}{d\tau} + \frac{1}{\tilde{r}} \, \frac{d\tilde{r}}{d\tau}\right) \, .
\end{align}
Due to the azimuthal symmetry, the following quantity is conserved~\cite{Sultana:2005tp}:
\begin{align}
\label{eq:tilde-L}
p_{3} \equiv \tilde{L} = \tilde{\eta}^4 \tilde{r}^2 \, \frac{d\phi}{d\tau} \, .
\end{align}
In Sec. \eqref{sec:Def-Angle}, we show that the dispersion relation depends on the value of $b = \tilde{L}/{\tilde{E}}$, which is dimensionless impact parameter. 
In the rest of this work, we obtain the change in photon propagation in the Sultana-Dyer space-time due to non-minimal coupling~\eqref{eq:S_Balakin}. We also compare our results with Schwarzschild space-time.

\section{Modified dispersion relations}
\label{sec:Dispersionrelations}

In this section,  from the equation of motion \eqref{eq:Final-EoM-withRicci},  we obtain the modified photon dispersion relation for the non-minimally coupled electromagnetic fields. We also show that the modification in the dispersion relation will be significant near strongly gravitating objects like black holes. Interestingly,  we show that the nature of the dispersion relation is different for time-independent (like Schwarzschild, Kerr) and time-dependent black hole (like Sultana-Dyer) spacetimes.

To derive the photon dispersion relation, we fix the photon momentum:  $p^{(\mu)} = (p^{(0)}, p^{(1)}, 0, \\ p^{(3)})$ --- confined in the orbital plane of the gravitating object. We choose a more generic electric field form than Ref.~\cite{Prasanna:2003ix}. Specifically,  we choose $E^{(i)} = \left[E^{(1)}, E^{(2)},\, E^{(3)}\right]$.

\subsection{Time-independent black hole spacetimes}
\label{sec:DRelation-Sch}

For the Schwarzschild spacetime:
\begin{align}
\label{eq:Schwarz}
ds^{2} =  - \left(1-\frac{\rH}{r} \right)\ dt^2  + 
 \left(1 -\frac{\rH}{r} \right)^{-1} dr^{2}\ + \ r^{2} \ d\Omega^{2} \, ,
\end{align}
using dimensionless coordinates $\tilde{r} = r/\rH, \, \tilde{t} = c\,t/\rH$  we obtain,
\begin{align}
ds^{2} = \rH^2 \, d\tilde{s}_{\rm Sch}^{2}
\end{align}
where,
\begin{align}
\label{def:dimensionless-Sch-metric}
 d\tilde{s}_{\rm Sch}^{2} = \left[- \left(1- \frac{1}{\tilde{r}} \right) \, d\tilde{t}^2 + \left(1 - \frac{1}{\tilde{r}}\right)^{-1} \, d\tilde{r}^2 + \tilde{r}^{2} d\Omega^2 \right] \, .
\end{align}
Since the line-element is independent of $\tilde{t}$ and $\phi$, we have the following invariant quantities (w.r.t. $\tau$):
\begin{align}
\label{def:tilde-E-L-Sch}
\tilde{E}_{\rm Sch} = - \left(1- \frac{1}{\tilde{r}} \right) \frac{d\tilde{t}}{d\tau} \, ; \,
\tilde{L}_{\rm Sch} = \tilde{r}^2 \frac{d\phi}{d\tau} \, .
\end{align}

The equation of motion~\eqref{eq:Final-EoM-withRicci} for the electric field $E^{(i)} = \left[E^{(1)}, E^{(2)},\, E^{(3)}\right]$ in the line-element~\eqref{def:dimensionless-Sch-metric} are:
\begin{align}
\label{eq:matrix_dispersion-Sch}
 \epsilon^{(i)}\,_{(j)} \, E^{(j)} = 0 \, .
\end{align}
where, the polarization tensor $\epsilon^{(i)}\,_{(j)}$ is, 
{
\tiny
\begin{align}
\epsilon^{(i)}\,_{(j)} = 
\begin{bmatrix}
-p_{(0)}^2+p_{(1)}^2 + p_{(3)}^2+\frac{2\left(2 p_{(0)}^2-2 p_{(1)}^2 +p_{(3)}^2\right) \lambda}{r^3} & 0 & 0 \\
0 & -p_{(0)}^2+p_{(1)}^2+p_{(3)}^2+\frac{2\left(-p_{(0)}^2+p_{(1)}^2-2 p_{(3)}^2\right) \lambda}{r^3} & 0 \\
-\frac{6p_{(1)} p_{(3)} \lambda}{r^3} & 0 & -p_{(0)}^2+p_{(1)}^2+ p_{(3)}^2 + \frac{2\left(-p_{(0)}^2+p_{(1)}^2+p_{(3)}^2\right) \lambda}{r^3}
\end{bmatrix}
\end{align}
}
{For the Schwarzschild spacetime, the Ricci tensor is zero; hence, those terms in the LHS of Eq. \eqref{eq:Final-EoM-withRicci} vanish. In order for the above Eq.~\eqref{eq:matrix_dispersion-Sch} to satisfy for 
arbitrary values of $[E^{(1)},\, E^{(2)},\,E^{(3)}]$,  the 
determinant of the polarization tensor  $\epsilon^{(i)}\,_{(j)}$ must vanish.  Demanding this leads to the following dispersion relation 
\begin{align}
\label{eq:dispersion_relation_sch}
& \qquad \qquad p^2 = \mathcal{C}_{\pm} \hspace{0.1cm}   p_{(3)}^2, \quad p^2 = 0\\
\label{def:coupling_pm_Sch} 
&\text{where,} \hspace{0.5cm} 
\mathcal{C}_{+}= \frac{6\tilde{\lambda}}{2\,\tilde{\lambda} + \tilde{r}^3 }; 
\hspace{0.5cm} \mathcal{C}_{-} =  \frac{6\tilde{\lambda}}{4\,\tilde{\lambda} -\tilde{r}^3  }; \qquad 
\tilde{\lambda} \equiv \lambda/r_{\rm H}^2 \, . 
\end{align}
Note that $\tilde{\lambda}$ is the dimensionless NMC parameter. It is important to note that the above dispersion relation differs from the one derived in Ref.~\cite{Prasanna:2003ix}. From the above expression~\eqref{eq:dispersion_relation_sch}, we infer that one of the dispersion relations is the same as the dispersion relation for minimally coupled photons. The modified photon dispersion relations are quadratic with $\tilde{r}, \tilde{\lambda}$ or, $r,\, \lambda-$dependent corrections. Interestingly, the photon dispersion relation in the tetrad frame constructed in Kerr spacetime (photons confined in the equatorial plane) is identical to Eq.~\eqref{eq:dispersion_relation_sch}. Appendix (\ref{sec:Dispersion relation with Kerr metric}) contains the detailed calculation for the Kerr spacetime in the orthonormal basis~\cite{Visser:2007fj}. Thus, the modified dispersion relation remains quadratic for the time-independent black hole spacetimes, like Schwarzschild and Kerr. In Ref.~ \cite{2023-Johnson.Jana.Shanki-gr-qc},  we obtained the explicit form of the modified dispersion relation in a slowly rotating Kerr BH. There we discussed the shadow formation by NMC photons in Schwarzschild and slowly rotating Kerr BH and their impact on BH observations. 
\subsection{Sultana-Dyer black hole}
\label{sec:DRelation-SD}

In this section, we show that the modified dispersion relation of the photons does not remain quadratic in the Sultana-Dyer black hole. We show it provides a key distinguishing feature between time-independent and Sultana-Dyer black hole spacetimes. 

To do that, we construct a local inertial frame corresponding to the Sultana-Dyer metric \eqref{eq:dimensionless-SDyer-metric}. 
Appendix~(\ref{sec:Tetrads Riemann in the local frame}) contains the details of the tetrads and the components of the Riemann tensor in the local inertial frame. Unlike the Schwarzschild and Kerr black holes, the Ricci tensor is non-zero for the Sultana-Dyer black hole. Hence, in principle, all terms of Eq. \eqref{eq:Final-EoM-withRicci} will contribute to the dispersion relation. 

Substituting the components of the Riemann tensor, Ricci tensor, and scalar in the local frame, as mentioned in \eqref{eq:SD-Riemann} into Eq.~\eqref{eq:Final-EoM-withRicci}, the equations of motion for the electric field $E^{(i)} = \left[E^{(1)}, E^{(2)},\, E^{(3)}\right]$ are:
\begin{align}
\label{eq:matrix_dispersion}
 \begin{bmatrix}
    M_{11} & 0 & M_{13} \\
    0 & M_{22} & 0 \\
    M_{31} & 0 & M_{33}
  \end{bmatrix}
  \begin{bmatrix}
    E^{(1)} \\
    E^{(2)} \\  
     E^{(3)}
  \end{bmatrix} 
  = 0
\end{align}
where, 
{
\footnotesize
\begin{align}
\nonumber
M_{11} &= p_{(0)}^2 \left(\frac{4 \tilde{\eta}_{0}^4 \tilde{\lambda}}{\tilde{\eta}^4 \tilde{r}^3}-\frac{8 \tilde{\eta}_{0}^4 \tilde{\lambda} (3 \tilde{r}+5)}{\tilde{\eta}^5 \tilde{r}^2 (\tilde{r}+1)}+\frac{88 \tilde{\eta}_{0}^4 \tilde{\lambda} (\tilde{r}+1)}{\tilde{\eta}^6 \tilde{r}}-1\right)+ \frac{p_{(1)} p_{(3)}^2}{p_{(0)}}\frac{12 \tilde{\eta}_{0}^4 \tilde{\lambda} }{\tilde{\eta}^5 \tilde{r} (\tilde{r}+1) } - p_{(0)} p_{(1)} \frac{32 \tilde{\eta}_{0}^4 \tilde{\lambda} }{\tilde{\eta}^5 \tilde{r} (\tilde{r}+1)} \\
& +p_{(1)}^2 \left(-\frac{4 \tilde{\eta}_{0}^4 \tilde{\lambda}}{\tilde{\eta}^4 \tilde{r}^3}+\frac{8 \tilde{\eta}_{0}^4 \tilde{\lambda} (7 \tilde{r}+5)}{\tilde{\eta}^5 \tilde{r}^2 (\tilde{r}+1)}+\frac{8 \tilde{\eta}_{0}^4 \tilde{\lambda} (\tilde{r}+1)}{\tilde{\eta}^6 \tilde{r}}+1\right) + p_{(3)}^2 \left(\frac{2 \tilde{\eta}_{0}^4 \tilde{\lambda}}{\tilde{\eta}^4 \tilde{r}^3}+\frac{4 \tilde{\eta}_{0}^4 \tilde{\lambda} (4 \tilde{r}+7)}{\tilde{\eta}^5 \tilde{r}^2 (\tilde{r}+1)}-\frac{64 \tilde{\eta}_{0}^4 \tilde{\lambda} (\tilde{r}+1)}{\tilde{\eta}^6 \tilde{r}}+1\right) \\
 M_{13} & = - \frac{p_{(1)}^2 p_{(3)}}{p_{(0)}} \, \frac{12 ^4 \tilde{\lambda} }{\tilde{\eta}^5 \tilde{r} (\tilde{r}+1) } - p_{(0)} p_{(3)} \, \frac{12 \tilde{\eta}_{0}^4 \tilde{\lambda} }{\tilde{\eta}^5 \tilde{r} (\tilde{r}+1)} + p_{(1)} p_{(3)} \, \frac{12 \tilde{\eta}_{0}^4 \tilde{\lambda}  \left(2 \tilde{\eta}+6 (\tilde{r}+1)^2\right)}{\tilde{\eta}^6 \tilde{r} (\tilde{r}+1)}\\
 M_{22} & = p_{(0)}^2 \left(-\frac{2 \tilde{\eta}_{0}^4 \tilde{\lambda}}{\tilde{\eta}^4 \tilde{r}^3}-\frac{4 \tilde{\eta}_{0}^4 \tilde{\lambda} (18 \tilde{r}+19)}{\tilde{\eta}^5 \tilde{r}^2 (\tilde{r}+1)}+\frac{88 \tilde{\eta}_{0}^4 \tilde{\lambda} (\tilde{r}+1)}{\tilde{\eta}^6 \tilde{r}}-1\right) 
 -  p_{(0)} p_{(1)} \frac{8 \tilde{\eta}_{0}^4 \tilde{\lambda}}{\tilde{\eta}^5 \tilde{r} (\tilde{r}+1)} \nonumber\\
& +p_{(1)}^2 \left(\frac{2 \tilde{\eta}_{0}^4 \tilde{\lambda}}{\tilde{\eta}^4 \tilde{r}^3}+\frac{4 \tilde{\eta}_{0}^4 \tilde{\lambda} (20 \tilde{r}+19)}{\tilde{\eta}^5 \tilde{r}^2 (\tilde{r}+1)}-\frac{64 \tilde{\eta}_{0}^4 \tilde{\lambda} (\tilde{r}+1)}{\tilde{\eta}^6 \tilde{r}}+1\right) + p_{(3)}^2 \left(-\frac{4 \tilde{\eta}_{0}^4 \tilde{\lambda}}{\tilde{\eta}^4 \tilde{r}^3}+\frac{64 \tilde{\eta}_{0}^4 \tilde{\lambda}}{\tilde{\eta}^5 \tilde{r}^2}-\frac{64 \tilde{\eta}_{0}^4 \tilde{\lambda} (\tilde{r}+1)}{\tilde{\eta}^6 \tilde{r}}+1\right)\\
M_{31} & = \frac{12 \tilde{\eta}_{0}^4 \tilde{\lambda} p_{(3)}^3}{\tilde{\eta}^5 \tilde{r} (\tilde{r}+1) p_{(0)}} - \frac{12 \tilde{\eta}_{0}^4 \tilde{\lambda} p_{(0)} p_{(3)}}{\tilde{\eta}^5 \tilde{r} (\tilde{r}+1)}+\frac{6 \tilde{\eta}_{0}^4 \tilde{\lambda} p_{(1)} p_{(3)} \left(\tilde{\eta}^2 (-\tilde{r}-1)-2 \tilde{\eta} \tilde{r} (2 \tilde{r}+3)+12 \left(\tilde{r}^2+\tilde{r}\right)^2\right)}{\tilde{\eta}^6 \tilde{r}^3 (\tilde{r}+1)} \\
M_{33} &= p_{(0)}^2 \left(-\frac{2 \tilde{\eta}_{0}^4 \tilde{\lambda}}{\tilde{\eta}^4 \tilde{r}^3}-\frac{4 \tilde{\eta}_{0}^4 \tilde{\lambda} (18 \tilde{r}+19)}{\tilde{\eta}^5 \tilde{r}^2 (\tilde{r}+1)}+\frac{88 \tilde{\eta}_{0}^4 \tilde{\lambda} (\tilde{r}+1)}{\tilde{\eta}^6 \tilde{r}}-1\right) 
 - \frac{12 \tilde{\eta}_{0}^4 \tilde{\lambda} p_{(1)} p_{(3)}^2}{\tilde{\eta}^5 \tilde{r} (\tilde{r}+1) p_{(0)}} - \frac{8 \tilde{\eta}_{0}^4 \tilde{\lambda} p_{(0)} p_{(1)}}{\tilde{\eta}^5 \tilde{r} (\tilde{r}+1)} \nonumber \\
&+p_{(1)}^2 \left(\frac{2 \tilde{\eta}_{0}^4 \tilde{\lambda}}{\tilde{\eta}^4 \tilde{r}^3}+\frac{4 \tilde{\eta}_{0}^4 \tilde{\lambda} (20 \tilde{r}+19)}{\tilde{\eta}^5 \tilde{r}^2 (\tilde{r}+1)}-\frac{64 \tilde{\eta}_{0}^4 \tilde{\lambda} (\tilde{r}+1)}{\tilde{\eta}^6 \tilde{r}}+1\right)  +p_{(3)}^2 \left(\frac{2 \tilde{\eta}_{0}^4 \tilde{\lambda}}{\tilde{\eta}^4 \tilde{r}^3}+\frac{4 \tilde{\eta}_{0}^4 \tilde{\lambda} (10 \tilde{r}+7)}{\tilde{\eta}^5 \tilde{r}^2 (\tilde{r}+1)}+\frac{8 \tilde{\eta}_{0}^4 \tilde{\lambda} (\tilde{r}+1)}{\tilde{\eta}^6 \tilde{r}}+1\right)
\end{align}
}
To satisfy Eq.~\eqref{eq:matrix_dispersion} for arbitrary values of $[E^{(1)}, E^{(2)},\, E^{(3)}]$ the determinant of the first matrix at the L.H.S of Eq.~\eqref{eq:matrix_dispersion} has to be zero, which is nothing but the dispersion relation. The SD BH is a time-dependent BH, so the dispersion relations differ from the Schwarzschild BH. Specifically, we have two sets of dispersion relations --- one quadratic and another quartic.

\subsubsection{Quadratic dispersion relation in SD BH (SD-I):}

The quadratic dispersion relation, which we refer as SD-I, in the local inertial frame is 
\begin{align}
\label{eq:SD-DRelation-Local1}
  f_{1}\, p_{(0)}^2 + f_{2}\, p_{(1)}^2 + f_{3}\, p_{(3)}^2 + f_{4}\, p_{(1)} p_{(0)}  &= 0 
\end{align}
where,
\begin{subequations}
\label{eq:SD-def-f1-5}
\begin{align}
f_{1}(\tilde{r},\tilde{\eta}) &= -\tilde{r}^3\, \tilde{\eta}^6 \left(\tilde{r}+1\right) + \tilde{\lambda} \, \tilde{\eta}_{0}^4  \, \left[88  \tilde{r}^2 (\tilde{r}+1)^2 - 2 \tilde{\eta}^2  (\tilde{r}+1)-4 \tilde{\eta}   \tilde{r} (18 \tilde{r}+19) \right]
\\
f_{2}(\tilde{r},\tilde{\eta}) & =  \tilde{\eta}^6 \tilde{r}^3 (\tilde{r}+1) + \tilde{\lambda}\,  \tilde{\eta}_{0}^4 \, \left[ 2 \tilde{\eta}^2 (\tilde{r}+1) + 4 \tilde{\eta}   \tilde{r} (20 \tilde{r}+19) - 64  \tilde{r}^2 (\tilde{r}+1)^2 \right]
\\
f_{3}(\tilde{r},\tilde{\eta}) &= \tilde{\eta}^6 \tilde{r}^3 (\tilde{r}+1) + \tilde{\lambda}\, \tilde{\eta}_{0}^4 \, \left[ -4 \tilde{\eta}^2   (\tilde{r}+1) + 64 \tilde{\eta}   \tilde{r} (\tilde{r}+1) - 64  \tilde{r}^2 (\tilde{r}+1)^2  \right]
\\
f_{4}(\tilde{r},\tilde{\eta}) &= - \tilde{\lambda} \,\tilde{\eta}_{0}^4 \left( 8 \tilde{\eta}  \tilde{r}^2  \right)
\end{align}
\end{subequations}
Using the tetrad relations in Eq. \eqref{eq:SD-tetrad}, we obtain the following dispersion relation in the non-inertial frame:
\begin{align}
\label{eq:SD-DRelation-NLocal1}
f_{1}^{\prime}\, p_{0}^2 + f_{2}^{\prime}\, p_{1}^2 + f_{3}^{\prime}\, p_{3}^2 + f_{4}^{\prime}\, p_{1} p_{0}  &= 0 
\end{align}
where,
\begin{subequations}
\label{eq:f1-f4-prime}
\begin{align}
f_{1}^{\prime} &=
\frac{\tilde{\eta}_{0}^4 (\tilde{r}+1)}{\tilde{\eta}^4 \tilde{r}} \left[-\tilde{\eta}^6 \tilde{r}^3 (\tilde{r}+1) + 2 \tilde{\lambda}\,\tilde{\eta}_{0}^4 \, \left(-\tilde{\eta}^2+(44-36 \tilde{\eta}) \tilde{r}^2-\tilde{\eta} (\tilde{\eta}+38) \tilde{r}+44 \tilde{r}^4+88 \tilde{r}^3\right) \right]
\\
f_{2}^{\prime} &= \frac{\tilde{\eta}_{0}^4 (\tilde{r}+1)}{\tilde{\eta}^4 \tilde{r}} \left[\tilde{\eta}^6 (\tilde{r}-1) \tilde{r}^3  - 2 \tilde{\lambda}\, \tilde{\eta}_{0}^4\,  \left(\tilde{\eta}^2-4 (10 \tilde{\eta}+11) \tilde{r}^2-(\tilde{\eta}-38) \tilde{\eta} \tilde{r}+32 \tilde{r}^4\right) \right]
\\
f_{3}^{\prime} &=  \frac{\tilde{\eta}_{0}^4 (\tilde{r}+1)}{\tilde{\eta}^4 \tilde{r}^2} \left[ \tilde{\eta}^6 \tilde{r}^3 - 4 \tilde{\lambda}\, \tilde{\eta}_{0}^4 \, \left(\tilde{\eta}^2 - 16 \tilde{\eta} \tilde{r}+ 16 \tilde{r}^3 + 16 \tilde{r}^2\right) \right]
\\
f_{4}^{\prime} &=  \frac{2 \tilde{\eta}_{0}^4 (\tilde{r}+1)}{\tilde{\eta}^4 \tilde{r}} \left[\tilde{\eta}^6 \tilde{r}^3 +  2 \tilde{\lambda}\, \tilde{\eta}_{0}^4 \, \left(\tilde{\eta}^2-2 (\tilde{\eta}+22) \tilde{r}^2+38 \tilde{\eta} \tilde{r} - 44 \tilde{r}^3\right) \right]
\end{align}
\end{subequations}
\subsubsection{Quartic dispersion relation in SD BH (SD-II and SD-III):}

The quartic dispersion relation in the local inertial frame is: 

{\footnotesize
\begin{align}
\label{eq:SD-DRelation-Local2}
f_{5}\,  p_{(0)}^{4} - f_{6}\,  p_{(0)}p_{(1)} ^{3} -  f_{7} \, p_{(1)}p_{(0)}^{3} +  f_{8}\,  p_{(1)}^{2}p_{(0)}^{2} + f_{9}\, \left(-p_{(0)}^{2} + p_{(1)}^{2} + p_{(3)}^{2}\right) + f_{10}\, \left(-p_{(0)}^{2} + p_{(1)}^{2} + p_{(3)} ^{2}\right)^2 = 0
\end{align}
}
where, 
\begin{subequations}
\label{eq:f5-f10}
\begin{align}
\label{eq:f5}
f_{5} & = 32 \,\lambda^2 \, \tilde{\eta}_{0}^8  \tilde{r} \big(6 \tilde{r} (\tilde{r}+1)^2-\tilde{\eta}  (2 \tilde{r}+3)\big) \big(3 \tilde{\eta} ^2 (\tilde{r}+1)-2 \tilde{\eta}  \tilde{r} (2 \tilde{r}+3)+12 \tilde{r}^2 (\tilde{r}+1)^2\big) \\
\label{eq:f6}
f_{6} & = 64\, \lambda ^2 \, \tilde{\eta} ^2 \tilde{\eta}_{0}^8 \, \tilde{r}^2 \,  \left[4 \tilde{r} (4 \tilde{r}+3)-3 \tilde{\eta}  (\tilde{r}+1)\right]  \\
\label{eq:f7}
f_{7} & = 64 \, \lambda ^2 \,\tilde{\eta}  \tilde{\eta}_{0}^8  \tilde{r}^2 \, \left[3 \tilde{\eta} ^2 (\tilde{r}+1)-4 \tilde{\eta}  \tilde{r} (2 \tilde{r}+3)+24 \tilde{r}^2 (\tilde{r}+1)^2\right] \\
\label{eq:f8}
f_{8} & = 192\, \lambda ^2 \tilde{\eta}  \tilde{\eta}_{0}^8   \tilde{r} \,  \left[3 \tilde{\eta} ^2 (\tilde{r}+1)^2 - \tilde{\eta}  \tilde{r} (\tilde{r} (\tilde{r} (3 \tilde{r}+13)+21)+9)+4 \tilde{r}^2 (4 \tilde{r}+3) (\tilde{r}+1)^2\right] \\
\label{eq:f9}
f_{9} & = 2 \, \lambda \, \tilde{\eta}_{0}^4  (\tilde{r}+1) p_{(0)}^2 \, \bigg[ 2 \tilde{\eta}_{0}^4 \lambda  \big(16 \tilde{\eta}  (\tilde{r}+1) (70 \tilde{r}+99) \tilde{r}^3+4 \tilde{\eta} ^2 (\tilde{r} (18 \tilde{r}-7)-87) \tilde{r}^2  
  \nonumber \\
 & +4 \tilde{\eta} ^3 (10 \tilde{r}+3) \tilde{r} + 3 \tilde{\eta} ^4 (\tilde{r}+1)-1488 (\tilde{r}+1)^3 \tilde{r}^4\big)+\tilde{\eta} ^6 \tilde{r}^3 \big(3 \tilde{\eta} ^2 (\tilde{r}+1)-10 \tilde{\eta}  \tilde{r} (2 \tilde{r}+3)\nonumber \\ 
 & +60 \tilde{r}^2 (\tilde{r}+1)^2\big)\bigg] - 8 \, \lambda \, \tilde{\eta}  \tilde{\eta}_{0}^4  p_{(0)} p_{(1)} \tilde{r}^2 (\tilde{r}+1)\, \bigg[5 \tilde{\eta} ^6 \tilde{r}^3-4 \tilde{\eta}_{0}^4 \lambda  \big(-7 \tilde{\eta} ^2-62 \tilde{\eta}  \tilde{r}+62 (\tilde{r}+1) \tilde{r}^2\big)\bigg] \nonumber \\ 
&  + 2 \, \lambda \tilde{\eta}  \tilde{\eta}_{0}^4  p_{(1)}^2 (\tilde{r}+1)\,  \bigg[ 2 \tilde{\eta}_{0}^4 \lambda  \big(4 \tilde{\eta}  (\tilde{r} (24 \tilde{r}+215)+129) \tilde{r}^2+4 \tilde{\eta} ^2 (4 \tilde{r}-3) \tilde{r}-3 \tilde{\eta} ^3 (\tilde{r}+1) \nonumber \\
& -32 (\tilde{r}+1) (4 \tilde{r}+3) \tilde{r}^3\big) +\tilde{\eta} ^6 \tilde{r}^3 (10 \tilde{r} (4 \tilde{r}+3)-3 \tilde{\eta}  (\tilde{r}+1))\bigg]\\
\label{eq:f10}
f_{10} & =  (\tilde{r}+1)^2 \bigg[ \tilde{\eta} ^{12} \tilde{r}^6 + 4 \, \lambda \tilde{\eta} ^6 \tilde{\eta}_{0}^4  \tilde{r}^3 \big(\tilde{\eta} ^2+14 \tilde{\eta}  \tilde{r}-14 (\tilde{r}+1) \tilde{r}^2\big) \nonumber \\
& -4 \tilde{\eta}_{0}^8 \lambda ^2 \big(-\tilde{\eta} ^4+8 \tilde{\eta}  (76 \tilde{r}+49) \tilde{r}^3+28 \tilde{\eta} ^2 (\tilde{r} - 6) \tilde{r}^2 - 28 \tilde{\eta} ^3 \tilde{r} + 128 (\tilde{r}+1)^2 \tilde{r}^4\big) \bigg]
\end{align}
\end{subequations}
It is important to note that $f_{5}-f_{8}$ and $f_{10}$ does not include any component of momentum, but $f_{9}$ does as all the terms in Eq.~\eqref{eq:SD-DRelation-Local2} are quartic in $p_{(\mu)}$. As the dispersion relation in the local frame is very complicated, the quartic dispersion in the non-inertial frame is mathematically more complicated. To proceed with the rest of the mathematical derivation, we exclude all the terms with $\tilde{\lambda}^2$ in Eq.~\eqref{eq:SD-DRelation-Local2}. This assumption is justified as we will show later that $\tilde{\lambda}<1$. Omitting terms with $\tilde{\lambda}^2$ in Eq.~\eqref{eq:SD-DRelation-Local2}, we split the quartic dispersion relation and obtain two quadratic dispersion relations in the local inertial frame as follows,
\begin{align}
\label{eq:SD-DRelation-Local2-part1} 
& p_{(0)}^2 \left(\frac{24 \tilde{\lambda} (r+1)}{\tilde{\eta}_{0}^2 r}-\frac{12 \tilde{\lambda}}{\tilde{\eta}_{0} r^2 (r+1)}-\frac{8 \tilde{\lambda}}{\tilde{\eta}_{0} r (r+1)}+\frac{6 \tilde{\lambda}}{r^3}-1\right)+ p_{(0)} p_{(1)} \, \frac{8 \tilde{\lambda}} {\tilde{\eta}_{0} r (r+1)} \nonumber \\ 
& +p_{(1)}^2 \left(\frac{12 \tilde{\lambda}}{\tilde{\eta}_{0} r^2 (r+1)}+\frac{16 \tilde{\lambda}}{\tilde{\eta}_{0} r (r+1)}-\frac{6 \tilde{\lambda}}{r^3}+1\right) + p_{(3)}^2  = 0 \\
\label{eq:SD-DRelation-Local2-part2}
& p_{(0)}^2 \left(\frac{96 \tilde{\lambda} (r+1)}{\tilde{\eta}_{0}^2 r}-\frac{48 \tilde{\lambda}}{\tilde{\eta}_{0} r^2 (r+1)}-\frac{32 \tilde{\lambda}}{\tilde{\eta}_{0} r (r+1)}-1\right) + p_{(0)} p_{(1)}\, \frac{32 \tilde{\lambda} }{\tilde{\eta}_{0} r (r+1)}\nonumber \\
& + p_{(1)}^2 \left(\frac{48 \tilde{\lambda}}{\tilde{\eta}_{0} r^2 (r+1)}+\frac{64 \tilde{\lambda}}{\tilde{\eta}_{0} r (r+1)}+1\right) + p_{(3)}^2 =  0
\end{align}
Using the tetrads defined in Appendix.~\eqref{sec:Tetrads Riemann in the local frame}, the above two dispersion relations in the non-inertial frame reduce to:
\begin{eqnarray}
\label{eq:SD-DRelation-NLocal2-part1}
a_{22} \, p_{1}^{2} + a_{12} \, p_{1}^{} + a_{02} &=& 0 \\
 \label{eq:SD-DRelation-NLocal2-part2}
a_{23} \, p_{1}^{2} + a_{13} \, p_{1}^{} + a_{03} &=& 0
\end{eqnarray}
where, 
{\footnotesize
\begin{subequations}
\begin{align}
\label{eq:a02}
a_{02} &= - 
\frac{\tilde{E}_{\rm SD}^2 \, \tilde{\eta}_{0}^2 }{\tilde{\eta}^4 \tilde{r}^4 } \, \left[ +\tilde{\eta}_{0}^2 (\tilde{r}+1) \tilde{r}^3 + \tilde{\lambda}\, \left(4 \tilde{\eta}_{0}  (2 r+3) r - 24  (r+1)^2 r^2 -6 \tilde{\eta}_{0}^2  (r+1)\right) \right]
+   \frac{\tilde{L}_{\rm SD}^2 \, \tilde{\eta}_{0}^4 }{\tilde{\eta}^4 \tilde{r}^2} \\
\label{eq:a12}
a_{12} &= - \frac{ 2\, \tilde{E}_{\rm SD}\, \tilde{\eta}_{0}^2 }{\tilde{\eta}^4 \tilde{r}^4 (r+1)} \left[r^4 \left(\tilde{\eta}_{0}^2 - 24 \tilde{\lambda}\right)+r^3 \left(\tilde{\eta}_{0}^2-48 \tilde{\lambda}\right) - 6 \tilde{\eta}_{0}^2 \tilde{\lambda}-6 (\tilde{\eta}_{0}-2) \tilde{\eta}_{0} \tilde{\lambda} r-4 \tilde{\lambda} r^2 (\tilde{\eta}_{0} (\tilde{r}-2)+6 \right] 
\\
\label{eq:a22}
a_{22} &= \frac{\tilde{\eta}_{0}^2}{\tilde{\eta}^4 \tilde{r}^4 (\tilde{r}+1)^2} \, \bigg[ \left(\tilde{\eta}_{0}^2 \left(\tilde{r}^2-1\right) r^4+\tilde{\eta}_{0}^2 \left(\tilde{r}^2-1\right) r^3\right) + \tilde{\lambda}\, \big(-6 \tilde{\eta}_{0}^2 \left(\tilde{r}^2-1\right)+8 r^2 \left(\tilde{\eta}_{0} \left(2 \tilde{r}^2+\tilde{r}-1\right)+3\right) \nonumber 
\\ & -6 (\tilde{\eta}_{0}-2) \tilde{\eta}_{0} \left(\tilde{r}^2-1\right) r+24 r^4+48 r^3\big)  \bigg] 
\\
\label{eq:a03}
a_{03} &= - \frac{ \tilde{E}_{\rm SD}^2 \, \tilde{\eta}_{0}^2}{\tilde{\eta}^4 \tilde{r}^3  } \left[ \tilde{\eta}0^2 \tilde{r}^2 (\tilde{r}+1)+16 \tilde{\lambda} \left(\tilde{\eta}_{0} (2 r+3)- 6 \tilde{r} (\tilde{r}+1)^2\right) \right] +  
\frac{\tilde{L}_{\rm SD}^2 \, \tilde{\eta}_{0}^4 }{\tilde{\eta}^4 \tilde{r}^2}
\\
\label{eq:a13}
a_{13} &= - \frac{2  \tilde{E}_{\rm SD}\,\tilde{\eta}_{0}^2  }{\tilde{\eta}^4 \tilde{r}^3 (\tilde{r}+1)}\, \left[ r^3 \left(\tilde{\eta}_{0}^2-96 \tilde{\lambda}\right)+r^2 \left(\tilde{\eta}_{0}^2-192 \tilde{\lambda}\right)+48 \tilde{\eta}_{0} \tilde{\lambda}-16 \tilde{\lambda} r (\tilde{\eta}_{0} (\tilde{r}-2) + 6  \right]
\\
\label{eq:a23}
a_{23} &= \frac{\tilde{\eta}_{0}^2 }{\tilde{\eta}^4 \tilde{r}^4 (\tilde{r}+1)^2} \bigg[ \left(\tilde{\eta}_{0}^2 \left(\tilde{r}^2-1\right) r^3+\tilde{\eta}_{0}^2 \left(\tilde{r}^2-1\right) r^2\right) + \tilde{\lambda}\, \big(32 r \left(\tilde{\eta}_{0} \left(2 \tilde{r}^2+\tilde{r}-1\right)+3\right)+48 \tilde{\eta}_{0} \left(\tilde{r}^2-1\right) \nonumber 
\\
& + 96 r^3+192 r^2\big)  \bigg]
\end{align}
\end{subequations}
}
$\tilde{E}_{\rm SD}$ and $\tilde{L}_{\rm SD}$ are related to the covariant form of momentum components as:
\begin{subequations}
\begin{align}
\label{eq:p0-SD}
{p_{0} = \frac{\tilde{\eta}^4}{\tilde{\eta}_{0}^4}\left(-   \, A \, \frac{d\tilde{\eta}}{d\tau} + C_{1} \, \frac{d\tilde{r}}{d\tau}\right) = - \tilde{E}_{\rm SD}}\\
\label{eq:p1-SD}
{p_{1} = \frac{C_{1}}{A} \tilde{E}_{\rm SD} +   \frac{\tilde{\eta}^4}{\tilde{\eta}_{0}^4}\left(\frac{C_{1}^2}{A} + B\right)\frac{d\tilde{r}}{d\tau}} \\
\label{eq:p3-SD}
{p_{3} = \frac{\tilde{\eta}^4}{\tilde{\eta}_{0}^4} C_{2} \, \frac{d\phi}{d\tau} = \tilde{L}_{\rm SD}}
\end{align}
\end{subequations} 
where, $\tilde{E}_{\rm SD},\, \tilde{L}_{\rm SD}$ are energy and angular momentum of the photon in the SD spacetime. Contravariant form of momentum $P^{\mu}$ is, $p^{0} = d\tilde{\eta}/d{\tau}, \, p^{1} = d\tilde{r}/d{\tau}, \,
p^{2} = d{\theta}/d{\tau}, \, p^{3} = d{\phi}/d{\tau}$;  $A, B, C_1, C_2$ are defined in Appendix~\eqref{eq:SD-metricdef} and $f_i, f_{i}^{\prime}$'s are defined above in Eqs. \eqref{eq:SD-def-f1-5}, \eqref{eq:f1-f4-prime} and \eqref{eq:f5-f10}. 

This is the first important key result that we would like to highlight. The dispersion relations~\eqref{eq:SD-DRelation-NLocal1},  \eqref{eq:SD-DRelation-NLocal2-part1} and \eqref{eq:SD-DRelation-NLocal2-part2},  for the SD spacetime,  are very different than the dispersion relations in Schwarzschild BH~\eqref{eq:dispersion_relation_sch}.  In the case of Schwarzschild,  the two dispersion relations~\eqref{eq:dispersion_relation_sch} are quadratic, and the third one ($p^{2}=0$)  is identical to the minimal coupling. However, in the SD spacetime, one of the dispersion relations is quadratic (SD-I), and another is quartic (SD-II and SD-III). The quadratic dispersion (SD-I) has a different form (compared to Schwarzschild) as it contains the cross-term $p_{0}p_{1}$. As mentioned earlier,  assuming  $\tilde{\lambda}$ to be small and ignoring  
$\tilde{\lambda}^2$ terms reduces the quartic dispersion to two quadratic dispersion relations~\eqref{eq:SD-DRelation-NLocal2-part1} (SD-II) and \eqref{eq:SD-DRelation-NLocal2-part2} (SD-III). Interestingly, both the quadratic dispersion relations contain the cross-terms $p_{0}p_{1}$. This indicates that the total deflection angle might differ for the two black hole spacetimes. 

In the next section, we calculate the expression for the deflection angle, followed by the total deflection angle of the photon near the black hole. We compare the time-independent (Schwarzschild) and time-dependent (SD) BH results. As the total deflection angle depends on $\tilde{\lambda}$, we also obtain a constraint on the NMC parameter.
\section{Deflection angle}
\label{sec:Def-Angle}

Potentially, the modification to the path taken by the photon leads to change in the following two observable quantities:
\begin{enumerate}
\item Photon arrival-time at the detector
\item Photon deflection-angle at the detector
\end{enumerate}
In Ref.~\cite{Prasanna:2003ix}, the authors used photon dispersion relation and computed the photon's arrival time in Schwarzschild spacetime. Since we consider time-dependent black hole spacetimes, the photon deflection angle is more suitable. Hence, in this work, we obtain the total deflection angle of the non-minimally coupled photon in both Schwarzschild and Sultana Dyer spacetimes and compare the results.

\subsection{Schwarzschild space-time}
\label{sec:DAngle-Sch}

To obtain the photon-deflection angle, we need to rewrite the dispersion relation \eqref{eq:dispersion_relation_sch} in the non-inertial frame. 
To do this, substituting the tetrads for the line-element~\eqref{def:dimensionless-Sch-metric} in the dispersion relation \eqref{eq:dispersion_relation_sch} leads to:
\begin{align}
\label{eq:dispersion-non-inertial-Sch}
\left(1 - \frac{1}{\tilde{r}}\right)\, p_{1}^{2} = \beta_{0}^{-2 }\left(1 - \frac{1}{\tilde{r}}\right)^{-1}\, p_{0}^{2} - \frac{1}{\tilde{r}^2} \, \left(1 -\mathcal{C}{\pm}\right) \, p_{3}^{2}
\end{align}
where, 
\begin{subequations}
\begin{align}
\label{eq:p0-p1-Sch}
p_{0} = - \left(1- \frac{1}{\tilde{r}} \right)  \frac{d\tilde{t}}{d\tau} := \tilde{E}_{\rm Sch}; &~~
p_{1} = \left(1 - \frac{1}{\tilde{r}}\right)^{-1} \frac{d\tilde{r}}{d\tau} \, ;\\
\label{eq:p3-Sch}
p_{3} = \tilde{r}^2 \frac{d\phi}{d\tau} := \tilde{L}_{\rm Sch}
\end{align}
\end{subequations}
Substituting $p_{0},\, p_{3}$ in Eq.~\eqref{eq:dispersion-non-inertial-Sch} and solving the resultant equation,  we obtain 
\begin{align}
\nonumber
\left(1 - \frac{1}{\tilde{r}}\right)\, p_{1}^2 = \left(1 - \frac{1}{\tilde{r}}\right)^{-1}\, \tilde{E}_{\rm Sch}^{2} - \frac{1}{\tilde{r}^2} \, \left(1 - \mathcal{C}_{\pm} \, \right) \, \tilde{L}_{\rm Sch}^{2} \, .
\end{align}
Substituting the form of $p_{1}$ from Eq.~\eqref{eq:p0-p1-Sch} in the above equation, we get:
\begin{align}
\label{eq:dtilder-dtau-Sch}
\frac{d\tilde{r}}{d\tau} = \tilde{L}_{\rm Sch} \left[\frac{1}{ b^2} - \frac{\left(1 - \frac{1}{\tilde{r}}\right)}{\tilde{r}^2} \left(1 - \mathcal{C}_{\pm} \right)\right]^{1/2}
\end{align}
where, $b \equiv \tilde{L}_{\rm Sch}/\tilde{E}_{\rm Sch}$ is the impact parameter.  {To obtain the deflection angle, we need to rewrite the above differential equation in terms of $\phi$. Thus,  dividing Eq.~\eqref{eq:dtilder-dtau-Sch} by $d\phi/d\tau = \tilde{L}_{\rm Sch}/\tilde{r}^2$,} the deflection angle 
$d\phi/d\tilde{r}$ is:
\begin{align}
\label{eq:deflection-angle-Sch}
\frac{d\phi}{d\tilde{r}} = \frac{1}{\tilde{r}^{2}} \ \left[\frac{1}{b^2} - \frac{\left(1 - \frac{1}{\tilde{r}}\right)}{\tilde{r}^2} \left(1 - \mathcal{C}_{\pm} \right)\right]^{-1/2} \, .
\end{align}
Note that the $-(+)$ sign corresponds to the deflection angle for the $+(-)$ polarization mode. {In the $\tilde{\lambda} = 0$ limit, the above expression reduces to the standard deflection angle for minimally coupled photons cf. Eq.~(6.3.35) in Ref.~\cite{1984-Wald-Book} provides a cut-off on the impact parameter for the photon. }

From the above expression, we see that the impact parameter ($b$) and NMC parameter ($\tilde{\lambda}$) can not be arbitrary for all values of $\tilde{r}$.   In the limit of $\tilde{r} \to \infty$, the second term inside the RHS bracket vanishes, suggesting no constraints on $b$ and $\tilde{\lambda}$. However,  close to the horizon, the second term inside the bracket in the RHS is finite. Thus, we see that the modification to the dispersion relation is the most significant close to the horizon and is negligible, very far from the horizon. This implies that  $b$ and $\tilde{\lambda}$ are constrained by the value of the horizon radius (which is in the dimensionless units in unity).  

To go about this, we need to determine
the radius of the photon sphere $r_{0}$ and its corresponding impact parameter $b_0$. At $r=r_{0}$ we have the following conditions:
\begin{align}
\label{eq:photonSphe-cond1and2}
\frac{d\tilde{r}}{d\tau}\biggr|_{r=r_{0}} =0;& \qquad 
\frac{d^2\tilde{r}}{d\tau^2}\biggr|_{r=r_{0}} = 0 
\end{align}
From Eq.~\eqref{eq:dtilder-dtau-Sch} and the conditions~\eqref{eq:photonSphe-cond1and2},  
we can obtain $\tilde{r}_0$ (or $r_0$) and the corresponding impact parameter $b_0$.  As mentioned above, the set $(\tilde{r}_0,\, b_{0})$ depends on the value of $\tilde{\lambda}$ as the photon dispersion relation depends on $\tilde{\lambda}$.  In a recent  work~\cite{2023-Johnson.Jana.Shanki-gr-qc},  
we showed that the values of $\tilde{r}_0,\, b_0$ increases with $\tilde{\lambda}$ for the dispersion relation~\eqref{eq:dispersion-non-inertial-Sch} with $\mathcal{C}_+$. Similarly, we showed that the value of $\tilde{r}_0,\, b_0$ decreases with $\tilde{\lambda}$ 
for the dispersion relation with $\mathcal{C}_-$.  In other words,  the value of $\tilde{r}_0,\, b_0$ depends on the type of polarization. To see this transparently,  let us return to the condition~\eqref{eq:photonSphe-cond1and2}. From the first condition,  the  impact parameter can be written as:
\begin{align}
\label{eq:b0-Sch}
b_{0} = \frac{\tilde{r}_{0}}{\sqrt{\left(1-\frac{1}{\tilde{r}}_{0}\right)(1-\mathcal{C}_{\pm})}}
\end{align}
Given that $b_{0}$ is a real positive number, we obtain the condition that the RHS of the above expression must be positive. This leads to the following constraint:
\begin{align}
\label{eq:lambda-constraint-Sch}
1-\mathcal{C}_{\pm} > 0
\end{align}
Using the definition of  $\mathcal{C}_{\pm}$ from Eq.~\eqref{def:coupling_pm_Sch} in the above expression, we get the following condition on $\tilde{\lambda}$: 
\begin{align}
\label{eq:Tilde-lambda-Constraint-Sch}
- 1.688 < \tilde{\lambda} < 0.844 \, .
\end{align}
It is important to note that the above condition is valid for both the dispersion relations. 

Having obtained the constraint on $\tilde{\lambda}$, we 
can obtain the deflection angles for the two polarization modes:
\begin{align}
\label{eq:deflection-angle-Sch1}
\frac{d\phi}{d\tilde{r}}\biggr|_{+}^{\rm Sch} &= \frac{1}{\tilde{r}^{2}} \ \left[\frac{1}{b^2} - \frac{\left(1 - \frac{1}{\tilde{r}}\right)}{\tilde{r}^2} \left(1 - \frac{6 \tilde{\lambda} }{2\tilde{\lambda}+\tilde{r}^3} \right)\right]^{-1/2} \, . \\
\label{eq:deflection-angle-Sch2}
\frac{d\phi}{d\tilde{r}}\biggr|_{-}^{\rm Sch} &= \frac{1}{\tilde{r}^{2}} \ \left[\frac{1}{b^2} - \frac{\left(1 - \frac{1}{\tilde{r}}\right)}{\tilde{r}^2} \left(1 -\frac{6 \tilde{\lambda} }{4\tilde{\lambda} - \tilde{r}^3} \right)\right]^{-1/2} \, . 
\end{align}
In the above two equations, the subscript $+, \, -$ corresponds to the deflection for the dispersion relation with $\mathcal{C}_+$ and $\mathcal{C}_-$ respectively. From the expression of $d \phi  / d\tilde{r}$, we evaluate the total deflection angle for the two dispersion relations,
\begin{align}
\label{eq:int-def-angle-Sch}
( \triangle \phi )^{\rm Sch} = 2\pi - \int^{\infty}_{\tilde{r}_0} \, d\tilde{r} \,\left(\frac{d\phi}{d\tilde{r}}\biggr|_{\pm}^{\rm Sch} \right)  
\end{align}
In the next section \eqref{sec:Comparison}, we obtain the total deflection angle for Schwarzschild and compare it with the total deflection angle for SD spacetime.

\subsection{Sultana-Dyer black hole}
\label{sec:DAngle-SD}
In Sec.~\eqref{sec:DRelation-SD}, we obtained two distinct --- quadratic (\ref{eq:SD-DRelation-Local1}) 
and quartic (\ref{eq:SD-DRelation-Local2}) --- dispersion relations for the NMC electromagnetic field in Sultana-Dyer spacetime. In the previous subsection,  for the Schwarzschild, $\tilde{\lambda}$,  we obtained constraint on $\lambda$ (see Eq.~\eqref{eq:Tilde-lambda-Constraint-Sch}).  In Appendix \eqref{sec:lambda from EM tensor}  demanding the energy density of NMC  EM fields or photons in Sultana-Dyer spacetime to be positive, we obtain a similar constraint on $\tilde{\lambda}$. Since these constraints are consistent for small values of 
$\tilde{\lambda}$,  assuming $\tilde{\lambda}$ to be small, we can rewrite the quartic dispersion relation \eqref{eq:SD-DRelation-Local2} as two quadratic relations~(\ref{eq:SD-DRelation-NLocal2-part1},\ref{eq:SD-DRelation-NLocal2-part2}).  In the rest of this section, we use these three dispersion relations to obtain the total deflection angle. 

To go about this, we first obtain a general expression for deflection angle in terms of $p_{1}$ and the conserved quantities $\tilde{E}_{\rm SD},\, \tilde{L}_{\rm SD}$. Rewriting the general expressions \eqref{eq:p1-SD}, \eqref{eq:p3-SD},  we have: 
\begin{align}
\label{eq:drdtau-SD}
\frac{d\tilde{r}}{d \tau} &= \frac{\frac{\tilde{\eta}_{0}^4}{\tilde{\eta}^4}\left(p_1 - \frac{C_1 \tilde{E}_{\rm SD}}{A}\right)}{\left(\frac{C_{1}^2}{A} + B\right)}  \\
\label{eq:dphidtau-SD}
\frac{d\phi}{d\tau} &= \left(\frac{\tilde{\eta}_{0}^4}{\tilde{\eta}^4}\right) \frac{\tilde{L}_{\rm SD}}{C_2} \, .
\end{align}
Dividing Eq.~\eqref{eq:dphidtau-SD} by Eq.~\eqref{eq:drdtau-SD} leads to the following general expression for deflection angle in SD spacetime: 
\begin{align}
\label{eq:dphidr}
\frac{d\phi}{d \tilde{r}} =  \frac{\tilde{L}_{SD}\left(\frac{C_{1}^2}{A} + B\right)}{C_{2}\left[p_1 - \frac{C_{1}}{A} \tilde{E}_{\rm SD}\right]} 
\end{align}
Note that $p_1$ in the above expression depends on 
the choice of the dispersion relation. In the previous section we obtained three dispersion relations (\ref{eq:SD-DRelation-NLocal1}, \ref{eq:SD-DRelation-NLocal2-part1} and \ref{eq:SD-DRelation-NLocal2-part2}).  Once we obtain $p_1$, we will obtain the total deflection angle. To go about that,  we first rewrite the quadratic dispersion relation \eqref{eq:SD-DRelation-NLocal1} in the non-inertial frame as:
\begin{align}
\label{eq:dispersion2}
a_{21} \, p_{1}^{2} + a_{11} \, p_{1}^{} + a_{01} = 0 \, ,
\end{align}
where,
\begin{align}
& a_{01}  = f_{1}\,^{\prime} \tilde{E}_{\rm SD}^2\ + f_{3}\,^{\prime}\tilde{L}_{\rm SD}^2;\quad 
a_{11}  = -f_{4}\,^{\prime} \tilde{E}_{\rm SD}\,, \quad
a_{21}   = f_{2}\,^{\prime}
\end{align}
Solving the above equation,  we obtain:
\begin{align}
\label{eq:p1-QuadraticDispersion}
p_{1} = \frac{-a_{11} \pm D_1}{2 a_{21}},~~~\mbox{where}~~~D_1 = \left(a_{11}^{2} - 4 a_{21}a_{01}\right)^{1/2} \, .
\end{align}
Substituting the above form of $p_1$ in Eq.\eqref{eq:dphidr}, we get:
\begin{align}
\label{eq:deflection-angleSD1}
\frac{d\phi}{d\tilde{r}}\biggr|_{\rm SD-I} &= \frac{\tilde{L}\left(\frac{C_{1}^2}{A} + B\right)}{C_{2}\left[\frac{-a_{11} \pm D_1}{2 a_{21}} - \frac{C_{1}}{A} \tilde{E}\right]}
\end{align}
Similarly,  from the quadratic dispersion relations~\eqref{eq:SD-DRelation-NLocal2-part1}, \eqref{eq:SD-DRelation-NLocal2-part2} we obtain $p_{1}$. Substituting those in Eq.\eqref{eq:dphidr}, we obtain: 
\begin{align}
\label{eq:deflection-angleSD2}
\frac{d\phi}{d\tilde{r}}\biggr|_{\rm SD-II} &= \frac{\tilde{L}\left(\frac{C_{1}^2}{A} + B\right)}{C_{2}\left[\frac{-a_{12} \pm D_2}{2 a_{22}} - \frac{C_{1}}{A} \tilde{E}\right]} \\
\label{eq:deflection-angleSD3}
\frac{d\phi}{d\tilde{r}}\biggr|_{\rm SD-III} &= \frac{\tilde{L}\left(\frac{C_{1}^2}{A} + B\right)}{C_{2}\left[\frac{-a_{13} \pm D_3}{2 a_{23}} - \frac{C_{1}}{A} \tilde{E}\right]} 
\end{align}
where, $D_2 = \left(a_{12}^{2} - 4 a_{22}a_{02}\right)^{1/2}$ and $D_3 = \left(a_{13}^{2} - 4 a_{23}a_{03}\right)^{1/2}$.
Note that here the subscripts SD-I, SD-II, SD-III in the LHS of the above equations correspond to the deflection angle for dispersion relations~(\ref{eq:SD-DRelation-NLocal1},\, \ref{eq:SD-DRelation-NLocal2-part1} and, \ref{eq:SD-DRelation-NLocal2-part2}) 
respectively. From Eqs.~(\ref{eq:deflection-angleSD1},\, \ref{eq:deflection-angleSD2} and, \ref{eq:deflection-angleSD3}) we can obtain the total deflection angle $\Delta \phi$, i. e., 
\begin{align}
\label{eq:totaldef-angle}
 ( \triangle \phi )^{\rm SD}  = 2\pi - \int^{\infty}_{\tilde{r}_0} d\tilde{r}\, \left(\frac{d\phi}{d\tilde{r}} \right)^{\rm SD} \, .
\end{align}
The above expression measures how much a photon has been deflected from a straight line while passing a compact object and reaching the observer at Earth. Physically, it informs us about the effect of the gravitational attraction of the compact object on the passing photon close to it.  

For a far away observer, it is \emph{only} possible to decipher whether the compact object is a Schwarzschild black hole or SD black hole by measuring the total deflection angle. In principle,  given the same black hole mass $M$ and NMC parameter $\tilde{\lambda}$,  the total deflection angle from these two black hole spacetimes must be different. As mentioned earlier,  starting from the energy-momentum tensor~\eqref{eq:NMC-EM-tensor},  we obtain the following constraint: 
\begin{equation}
-\frac{\rH^2}{2}  < \lambda < \rH^2 \quad 
\Longrightarrow \quad 
-\frac{1}{2} < \tilde{\lambda} < 1
\label{eq:constraint-tilde-lambda-SD}
\end{equation}
Details are in Appendix~\eqref{sec:lambda from EM tensor}. 

\begin{table}
\begin{center}
\begin{tabular}{|p{3.5cm}|p{3cm}|p{3cm}|p{3cm}|p{3cm}| }
 \hline
Black hole mass  & $\rH$ (cm) & $\lambda_{\rm SD}$  (cm$^2$)   & $\lambda_{\rm Sch}$  (cm$^2$)  \\
 \hline
 Supermassive & $\sim 10^{11}$ & $\sim 10^{22}$ & $\sim 10^{22}$ \\ 
 \hline
  Solar-mass  & $2.9 \times 10^{5}$ & $8.41 \times 10^{10}$ & $7.01 \times 10^{10}$ \\
 \hline
 $10^{-3}M_{\odot}$ & $2.9 \times 10^{2} $  
 & $8.41 \times 10^{4} $& $7.01 \times 10^{4}$ \\
 \hline
 $10^{-5}M_{\odot}$ & $2.9  $  
 & $8.41 $& $7.01 $ \\
 \hline
 $10^{-10}M_{\odot}$ & $2.9 \times 10^{-5} $  
 & $8.41\times 10^{-10} $& $7.01  \times 10^{-10}$ \\
 \hline
\end{tabular}
\end{center}
\caption{Upper limit on $\lambda$ for different mass ranges: $\lambda_{\rm SD}$ and $\lambda_{\rm Sch}$ correspond to $\lambda$ determined in Sultana-Dyer and Schwarzschild space-time respectively.}
\label{table}
\end{table}

\subsection{Bounds on \texorpdfstring{$\lambda$}{\lambda}}
This section compares the constraint on $\lambda$ we obtained with the literature. In Refs.~\cite{Drummond:1979pp,1994-LafranceRobertC-PRD,2014-Camanho-JHEP,2021-Bellazzini-JHEP,2020-deRham-PRD,2020-AccettulliHuber-PRD}, several authors have studied the effect of vacuum polarization on photon propagation in the gravitational field. In these, they obtained $\lambda \sim \alpha/(90\, \pi m_{e}^{2})$ by considering the case where the photon energies are small compared to the electron mass, i.e., $\hbar \omega \ll m_e c^2$ ($m_{e}$ is the mass of an electron). However, their analysis is valid up to Compton wavelength of electron $\lambda_{e}$, corresponding to photons of energy small compared to the electron mass. They showed that the constraint on the NMC parameter could be expressed as $\lambda \sim 1/\Lambda^2$(in $\hbar=c=1$ unit) where $\Lambda$ is the cutoff of the theory, and further, they have shown that $\lambda \sim 1/M_{\rm Pl}^2$ for a Planck sized black hole. Drummond and Hathrell's analysis only applies in weak gravity limit~\cite{Shore:2003zc,Donoghue:2015nba,Goon:2016une}.

Interestingly, the above constraint is consistent with the results derived in the previous subsection. We obtained $\lambda < \rH^2$ for both the black hole spacetimes. For a Planck mass BH $(M_{\rm Pl})$, the Schwarzschild radius can formally be written as $\rH^{\rm Pl} \sim 2/M_{\rm Pl}$. Thus, $\lambda < \rH^2$ leads to the same constraint $\lambda \sim 1/M_{\rm Pl}^2$ mentioned in literature.

Table.~\ref{table} lists the upper limit of $\lambda$ for different black hole masses. From the table, we infer that the smaller black hole mass will provide stringent constraints on the NMC parameter compared to supermassive black holes. This is consistent with the discussion in the Introduction. As mentioned in the Introduction, the square root of the Kretschmann scalar (${\cal K}$) measures the intensity of the gravitational field. From the Table.~\ref{table}, we see that the strong gravity regime provides a stringent constraint on the NMC parameter.
For instance, in Ref. \cite{2023-Jana.Shanki-MG16, 2023-Johnson.Jana.Shanki-gr-qc}, The observable photon ring does not form for ${\lambda}<-\,4.2 \times 10^{22}$\,cm$^2$  for $(+)$ mode and for ${\lambda} > 2.1 \times 10^{22}$\,cm$^2$ for the $(-)$ mode.

In Ref.~\cite{Prasanna:2003ix}, the authors obtained constraints on $\lambda$ considering photons in Schwarzschild spacetime and considering signals from radar ranging past the Sun, they found $\lambda \sim 1.1 \times 10^{20}~\rm{cm}^2$, which is about three orders of magnitude more stringent than the one obtained in Ref.~\cite{Bedran1986AnEO}. 
Thus, compared to Ref.~\cite{Prasanna:2003ix},  the bound obtained here is more stringent --- by over ten orders of magnitude. 
Note that, in the case of signals coming from binary pulsar PSR B1534+12, the bound on 
$\lambda \sim 0.6 \times 10^{11}\,$cm$^2$~\cite{Stairs:1999zr, Stairs:2002cw}. It matches the bound on $\lambda$ we obtained for the Schwarzschild and Sultana-Dyer black hole.

\section{Comparing total deflection angles in Schwarzschild and Sultana-Dyer black holes}
\label{sec:Comparison}

In this section, we present the final results of this work, comparing the total deflection angle of NMC photons near time-independent (Schwarzschild) and time-dependent(Sultana-Dyer) black holes.

As mentioned earlier,  the total deflection angle depends on the radius of the photon sphere $(r_0)$ and the NMC parameter $(\lambda$). To go about this, we demand the following conditions:
\begin{align}
\label{eq:photonSphe-cond1and2-a}
\frac{d\tilde{r}}{d\tau}\biggr|_{r=r_{0}} =0;& \qquad 
\frac{d^2\tilde{r}}{d\tau^2}\biggr|_{r=r_{0}} = 0 
\end{align}
For SD spacetime,  using the expression \eqref{eq:drdtau-SD}  and the above conditions, we can obtain $\tilde{r}_0 \equiv r_0/r_H$ (radius of the photon sphere scaled with the horizon radius $r_H$) and the corresponding impact parameter $b_0$. Unlike Schwarzschild, the three dispersion relations (\ref{eq:deflection-angleSD1}, 
\ref{eq:deflection-angleSD2}, \ref{eq:deflection-angleSD3}) for SD spacetime are more involved. Hence, it is not possible to obtain analytical expressions for the same. \ref{fig:r0-b0-plots}
contains $(\tilde{r}_0, \, b_0)$ for the two dispersion relations for Schwarzschild and three dispersion relations for SD.  
\begin{figure}[!htb]
\centering
\subfigure[\label{fig:1a}]{
\includegraphics[scale=0.3]{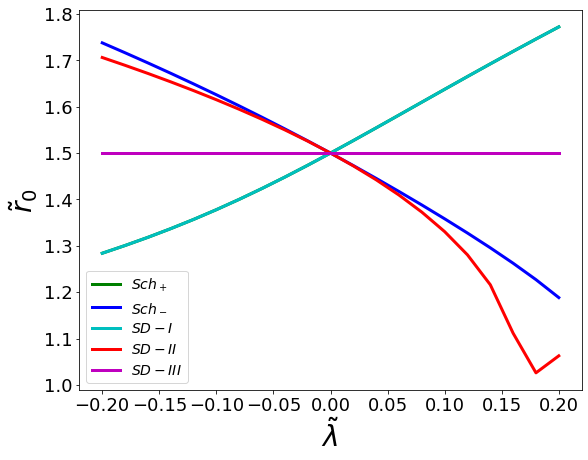}}
\quad
\subfigure[\label{fig:1b}]{
\includegraphics[scale=0.3]{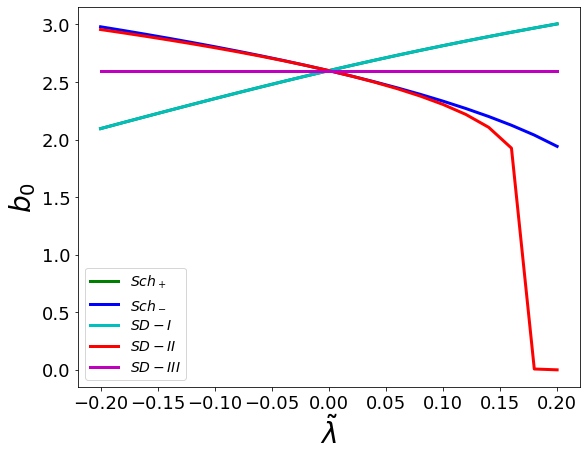}}
\caption{$\tilde{r}_0$ and corresponding $b_0$ as a function of $\tilde{\lambda}$ in Schwarzschild and Sultana Dyer spacetimes. The cyan, red, and magenta curves correspond to SD-I, SD-II, and SD-III dispersion relation results, respectively. The green and blue curves represent results for dispersion relation for (+) and (-) polarization modes in Schwarzschild spacetime, respectively.
}
\label{fig:r0-b0-plots}
\end{figure}

From \ref{fig:r0-b0-plots},  we infer the following : 
First, for the Dispersion relation-I (SD-I), which is the quadratic dispersion relation, the values of $\tilde{r}_0, \, b_0$ increases with the increase of $\tilde{\lambda}$. 
It is interesting to see that for $-0.2 \leq \tilde{\lambda} \leq 0.2$, $(\tilde{r}_0,\, b_0)$ for SD-I and $(+)$ polarization mode (Schwarzschild spacetime) are identical (The green and cyan curves overlap in Fig.~\ref{fig:1a},\ref{fig:1b}). However, $(\tilde{r}_0,\, b_0)$ differs for other dispersion relations in SD and Schwarzschild spacetime values.
Second, for dispersion relation SD-II, these values decrease as $\tilde{\lambda}$ increases. Interestingly, for $\tilde{\lambda} = 0.2$, $\tilde{r}_0$ is close to $1$, implying that the radius of the photon sphere is close to the horizon radius. {For the same $\tilde{\lambda}$, the impact parameter $b_0 = 0.0088$, which is less than the horizon radius, and implies that for $\tilde{\lambda} = 0.2$  no photon sphere can be formed for SD-II mode. We have verified that the photon sphere will not be formed for the SD-II mode when $\tilde{\lambda} \geq 0.168$. }
This is similar to the Schwarzschild spacetime. In this case, the photon sphere will not be formed at $\tilde{\lambda} \leq -0.5$ for the $\mathcal{C}_+$ mode and $\tilde{\lambda} \geq 0.25$ for the $\mathcal{C}_-$ mode of polarisations~\cite{2023-Johnson.Jana.Shanki-gr-qc}.
%
%
Lastly, for the Dispersion relation-III (SD-III), $\tilde{r}_0, \, b_0$ are constants irrespective of any value of $\tilde{\lambda}$.

\begin{figure}[!htb]
\centering
\subfigure[\label{fig:2a}]{
\includegraphics[scale=0.35]{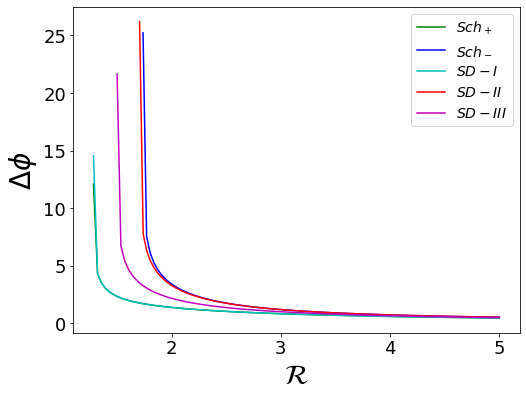}}
\quad
\subfigure[\label{fig:2b}]{
\includegraphics[scale=0.35]{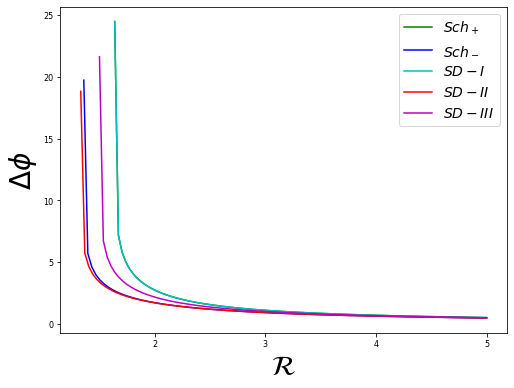}}
\caption{Total deflection angles near Schwarzschild and Sultana Dyer black holes for different dispersion relations at separate values of $\tilde{\lambda}$. Fig (a) corresponds to $\tilde{\lambda}=-0.2$ and (b) corresponds to $\tilde{\lambda}=0.1$. These results comprise the signatures of any solar-mass BH as per constraint~\eqref{eq:Tilde-lambda-Constraint-Sch} and~\eqref{eq:constraint-tilde-lambda-SD}.}
\label{fig:def-angle-plots}
\end{figure}
\begin{figure}[!htb]
\centering
\subfigure[\label{fig:3a}]{
\includegraphics[scale=0.35]{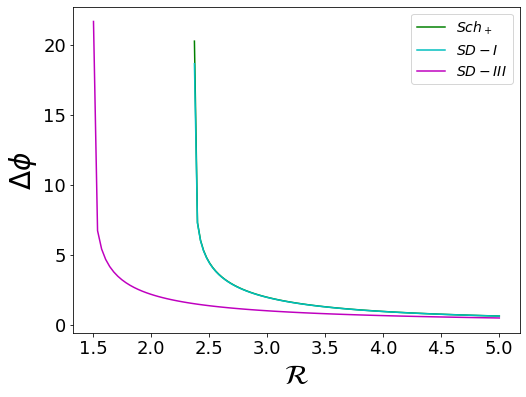}}
\quad
\subfigure[\label{fig:3b}]{
\includegraphics[scale=0.35]{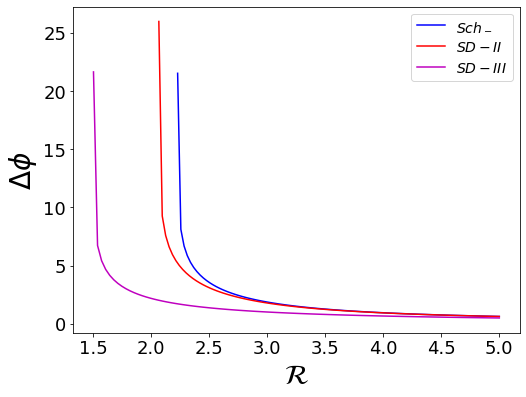}}
\caption{Total deflection angles near Schwarzschild and Sultana Dyer black holes for different dispersion relations at (a)$\tilde{\lambda}=0.8$ and (b)$\tilde{\lambda}=-0.8$ which corresponds to BH mass $M= 10^{-5} \times M_{\odot}$.}
\label{fig:def-angle-plots2}
\end{figure}

For the three dispersion relations of Sultana-Dyer and the two dispersion relations of Schwarzschild, we can calculate the total deflection angle $\Delta \phi$ numerically for different $\tilde{\lambda}$ values: 
\begin{align}
\label{eq:totaldef-anglefin}
 ( \triangle \phi )  = 2\pi - \int^{\mathcal{R}}_{\tilde{r}_0} d\tilde{r}\, \left(\frac{d\phi}{d\tilde{r}} \right) \, .
\end{align}
Here, $\tilde{r}_0$ is the radius of the photon sphere, scaled with BH horizon radius $\rH$, and ${\cal R}$ is in multiples of $\rH$. \ref{fig:def-angle-plots} and \ref{fig:def-angle-plots2} contain the total deflection angle as a function of ${\cal R}$ for all the dispersion relations and for different value of $\tilde{\lambda}$. While \ref{fig:def-angle-plots} is the plot of deflection angle with ${\cal R}$ for stellar-mass black holes, \ref{fig:def-angle-plots2} is the plot of deflection angle with ${\cal R}$ for PBHs of mass $10^{-5} M_{\odot}$. As mentioned earlier, the event horizon near PBH of this mass exhibits strong curvature.

From the figures, we infer the following: First,  as ${\cal R}$ increases,  $\Delta \phi$ saturates to a finite value. This is expected because the deflection angle must be large close to the horizon and will be negligible as we go away from the horizon. This trend is seen for all the dispersion relations for both the spacetimes. 
Second, for $\tilde{\lambda}=-0.2$, $\Delta \phi$ is highest for SD-II and lowest for the (+)-polarization mode in Schwarzschild BH.
Third,  for $\tilde{\lambda}=0.1$, $\Delta \phi$ is highest for both SD-I and the (+)-polarization mode in Schwarzschild spacetime. 

{From the table \eqref{table}, the upper limit of NMC parameter is $\sim 8$ for PBH of mass $M=10^{-5} M_{\odot}$. This implies $\tilde{\lambda} \sim 0.8$. \ref{fig:def-angle-plots2} contains the plot of the total deflection angle $\Delta \phi$ angle of NMC photons for Schwarzschild and SD spacetimes for this specific value of $\tilde{\lambda}$. For Schwarzschild, the photon sphere exists only for one mode ($\pm$ polarization for $\pm 0.8$). For SD-I and SD-III modes, the photon sphere exists for $\tilde{\lambda} = 0.8$, while for SD-II and SD-III modes, the photon sphere exists for $\tilde{\lambda} = -0.8$. Thus,  the total deflection angle $\Delta \phi$ of NMC photons near Schwarzschild and SD spacetime will have a contribution \emph{only} from the other modes. In other words, the photons forming the photon sphere are highly polarized, which can potentially lead to observable implications.
%

}

\section{Conclusions and Discussions}
\label{sec:conclusion}

In this work, we carried out a detailed analysis of the propagation of non-minimally coupled photons to gravity. To obtain a constraint on the coupling parameter, we calculated the deflection angle of a photon in the vicinity of a dynamical, spherically symmetric black hole described by the Sultana-Dyer spacetime. To our knowledge, the detailed evaluation of the deflection angle in a black-hole spacetime to constrain the non-minimal coupling parameter is new. 

Using the Eikonal approximation, we obtained the dispersion relations for the two polarization modes of electromagnetic fields in Schwarzschild and Sultana-Dyer spacetimes. We showed that the two dispersion relations corresponding to the two polarizations differ from Schwarzschild and Kerr [cf. Eq. \eqref{eq:dispersion_relation_sch}]. In the case of Schwarzschild and Kerr, the two dispersions are quadratic, and the modifications to the dispersion relation occur in the $p_{(3)}$ component of the momentum. However, in the case of Sultana-Dyer, the modification to the dispersion relation [cf. Eq. \eqref{eq:SD-DRelation-NLocal1}, \eqref{eq:SD-DRelation-NLocal2-part1}, \eqref{eq:SD-DRelation-NLocal2-part2}] occurs in all the components of the momentum. {The key difference between the dispersion relations in these two spacetimes is that for Schwarzschild, both are quadratic, whereas for Sultana-Dyer, one is quartic. In the small $\tilde{\lambda}$ limit, the quartic dispersion relation reduces to two quadratic dispersion relations .}

We compute the total deflection angle for non-minimally coupled photons near Schwarzschild and Sultana Dyer black holes and show that both these black holes have similar signatures regarding the total deflection angle and radius of the photon sphere. Interestingly, our analysis shows that PBH provides a stringent constraint on the NMC parameter. Thus, we conclude that irrespective of whether the PBH can be described as isolated or cosmological black holes, the constraint on the NMC parameter is similar. 

The constraint on $\lambda$ for the stellar mass black holes is better than that obtained by Prasanna and Mohanty~\cite{Prasanna:2003ix}. In Ref.~\cite{Prasanna:2003ix},  the authors used photon dispersion relation to compute the photon's arrival time and obtained $\lambda \sim 1.1 \times 10^{20}~\rm{cm}^2$. The constraint we have obtained is ten orders of magnitude more stringent compared to Ref. ~\cite{Prasanna:2003ix} and 13 orders of magnitude better than Ref.~\cite{Bedran1986AnEO}. Interestingly, the bound we have obtained matches with the bound obtained from the binary pulsar PSR B1534+12 is $\lambda \sim 0.6 \times 10^{11}\,$cm$^2$~\cite{Stairs:1999zr, Stairs:2002cw}. This suggests looking for constraints on $\lambda$ in the next-generation EHT/VLBI measurement~\cite{2019-Akiyama-Astrophys.J.Lett.,2022-Akiyama-Astrophys.J.Lett.}. With increased
sensitivity, {such as ngVLA}, $1~\mu as$ astrometry can be achieved, which can potentially detect deflection
angle from PBH of mass $10^{-5} M_{\odot}$ which are of the same size as Jovian-mass planets~\cite{2019-Reid-WhitePaper}.

In the current analysis, we have not considered the frequency dependence of the polarization modes. However, the frequency dependence of the deflection angle can provide a helpful way to test the non-minimal coupling strength and can potentially be measured in radio frequencies~\cite{Robishaw:2018ylp}. The entire analysis is focused on the spherically symmetric spacetimes, as PBH does not have spin. However, in the NS-NS and NS-BH mergers, the final black hole is expected to be axially symmetric (like Kerr). Therefore, we need to extend the analysis for the rotating black-hole spacetimes. This is currently under investigation.

\acknowledgments{We thank S. Mahesh Chandran and Ashu Kushwaha for their comments on the previous version of the manuscript. This work is supported by the SERB-CRG/2022/002348 grant.}




%


\appendix	 

\section{Photon dispersion relations in Kerr space-time}
\label{sec:Dispersion relation with Kerr metric}

The Kerr metric in Boyer–Lindquist coordinates is:
%


\begin{align}
\label{eq:Kerr-Boyer-Lindquist-coordinates}
\nonumber
ds^{2} & =  -\left[1- \frac{2Mr}{r^2 + a^2 \, \cos^{2}\theta}\right] \, dt^2 - \frac{4Mra \sin^2 \theta}{r^2 + a^2\, \cos^{2}\theta}\, dt \, d\phi \\ 
& + \left[\frac{r^2 + a^2\, \cos^{2}\theta}{r^2 - 2Mr + a^2}\right]\, dr^2 + \left(r^2 + a^2\, \cos^{2}\theta\right)\, d\theta^2 + \left[r^2 + a^2 + \frac{2Mra^2 \sin^2 \theta}{r^2 + a^2\, \cos^{2}\theta}\right]\, \sin^2 \theta \, d\phi^2 
\end{align}
%


In the orthogonal basis, the tetrads are of the form (see, for instance, Ref.~\cite{Visser:2007fj}):
\begin{align}
e^{(\mu)}\, _{\nu} & = 
\begin{bmatrix}
    \sqrt{\frac{r^2 - 2Mr + a^2}{r^2 + a^2\, \cos^{2}\theta}} & 0 & 0 & -a\sin^2\theta \sqrt{\frac{r^2 - 2Mr + a^2}{r^2 + a^2\, \cos^{2}\theta}}\\
    0 & \sqrt{\frac{r^2 + a^2\, \cos^{2}\theta}{r^2 - 2Mr + a^2}} & 0 & 0 \\
    0 & 0 & \sqrt{\frac{r^2 + a^2\, \cos^{2}\theta}{\sin^2 \theta}} & 0 \\
    -a\sqrt{\frac{\sin^2\theta}{r^2 + a^2}} & 0 & 0 & \sqrt{\sin^2 \theta (r^2 + a^2)}\\
\end{bmatrix}  \\[6pt]
e_{\nu} \,^{(\mu)}\, & = 
\begin{bmatrix}
   \frac{r^2 + a^2}{\sqrt{(r^2 - 2Mr + a^2)(r^2 + a^2\, \cos^{2}\theta)}} & 0 & 0 & \frac{a}{\sqrt{(r^2 - 2Mr + a^2)(r^2 + a^2\, \cos^{2}\theta)}}\\
    0 & \sqrt{\frac{r^2 + a^2\, \cos^{2}\theta}{r^2 - 2Mr + a^2}} & 0 & 0 \\
    0 & 0 & \sqrt{\frac{r^2 + a^2\, \cos^{2}\theta}{\sin^2 \theta}} & 0 \\
    a\sqrt{\frac{\sin^2\theta}{r^2 + a^2}} & 0 & 0 & \frac{1}{\sqrt{\sin^2 \theta (r^2 + a^2)}}\\
\end{bmatrix} 
\end{align}
For the equatorial ($\theta = \frac{\pi}{2}$) plane, the non-zero components of the Riemann tensor are:
\begin{align}
& R_{(0)(1)(0)(1)} =-2R_{(0)(2)(0)(2)} = -2 R_{(0)(3)(0)(3)} = 2R_{1)(2)(1)(2)} \nonumber \\
& = 2R_{(1)(3)(1)(3)} = -R_{(2)(3)(2)(3)} = \frac{2M}{r^3} \, .
\end{align}
The above components are identical to the components of (inertial frame) Riemann tensor in Schwarzschild geometry. Hence, we obtain the same dispersion relation as 
as in Eq.~\eqref{eq:dispersion_relation_sch}. In the non-inertial frame, the dispersion relation \eqref{eq:dispersion_relation_sch} leads to:
\begin{align}
\nonumber
& p_{1}^2\, \frac{r^2}{(r^2 - 2Mr + a^2)}  + p_{0}^{2}\left[-\frac{(r^2 + a^2)^2}{r^2(r^2 - 2Mr + a^2)} + \frac{a^2}{r^2 + a^2}\left(1 \mp 12 \lambda \frac{M}{r^3}\right)\right] \\
\nonumber
& + 2p_{0}p_{3} \left[-\frac{a(r^2 + a^2)}{r^2(r^2 - 2Mr + a^2)} + \frac{a}{r^2 + a^2}\left(1 \mp 12 \lambda \frac{M}{r^3}\right)\right] \\
\label{eq:dispersion-Kerr-non-inertial}
& + p_{3}^2 \left[-\frac{a^2}{r^2(r^2 - 2Mr + a^2)} + \frac{1}{r^2 +a^2}\left(1 \mp 12 \lambda \frac{M}{r^3}\right)\right] = 0
\end{align}
Since this is different from that of Schwarzschild \eqref{eq:dispersion-non-inertial-Sch}, this leads to a different time-of-arrival and deflection angle of photons compared to the Schwarzschild metric. As we show in Sec. \eqref{sec:DRelation-SD}, in the case of Sultana-Dyer black hole, the dispersion relation is no more quadratic.

\section{Sultana-Dyer black hole: Tetrads and Riemann tensor}
\label{sec:Tetrads Riemann in the local frame}

The local inertial frame for the Sultana-Dyer black hole 
\eqref{eq:dimensionless-SDyer-metric} is related to the non-inertial frame via the tetrads:
\begin{align}
\label{eq:relation-tetrad}
e^{(\mu)}\, _{\alpha} \, e^{(\nu)}\, _{\beta} \, g^{\alpha\beta} = \eta^{(\mu)(\nu)} 
\end{align} 
where, $\eta^{(\mu)(\nu)}$ is the metric for local Minkowski space-time and $g^{\alpha\beta}$ is metric for the non-inertial frame.
For the line-element \eqref{eq:dimensionless-SDyer-metric}, we have:
\begin{subequations}
\label{eq:SD-tetrad}
\begin{align}
\label{eq:tetrad1}
e^{0}\, _{(0)} = \frac{\sqrt{B}}{\tilde{\eta}^2} ;& \qquad 
e^{1}\, _{(0)} = -\frac{C_{1}}{\tilde{\eta}^2 \sqrt{B}} 
%
e^{1}\, _{(1)} = \frac{1}{\tilde{\eta}^2\sqrt{B}} ;  \\
\label{eq:tetrad3}
e^{2}\, _{(2)} = \frac{1}{\tilde{\eta}^2\sqrt{C_{2}} }; & \qquad
e^{3}\, _{(3)} = \frac{1}{\tilde{\eta}^2\sqrt{C_{2}} \sin \theta} 
\end{align}
\end{subequations}
where 
\label{eq:SD-metricdef}
\begin{align}
A(\tilde{r}) = 1 - \frac{1}{\tilde{r}} ;  \quad C_1(\tilde{r}) = \frac{ \beta_0 }{\tilde{r}}; \quad
B(\tilde{r})=  1 + \frac{1}{\tilde{r}} ;  \quad C_2(\tilde{r}) = \tilde{r}^2 
\end{align}
%
Components of inverse tetrad, $e_{\mu}\, ^{(\nu)}$ can also be calculated. The 
four-momentum of photon in non-inertial ($p_{\mu}$) and inertial ($p_{(\nu)}$) frames are related by $p_{\left(\mu\right)} = e^{\nu}\, _{\left(\mu\right)} \hspace{0.2cm} p_{\nu}$. 

We list below the non-zero components of the Riemann tensor in the inertial frame:
\begin{subequations}
\label{eq:SD-Riemann}
\begin{align}
\nonumber
& R_{(0)(1)(0)(1)} = -\, \frac{1}{\tilde{r}^3 (\tilde{r}+1)^3 \tilde{\eta}^6}  
[ \tilde{r}^2 \left(3 \beta_{0}^2 \tilde{\eta}^2+6 \beta_{0} \tilde{\eta}-2\right)+\beta_{0}^2 \tilde{\eta}^2+2 \tilde{r}^4 (\beta_{0} \tilde{\eta}-6)+\tilde{r}^3 \left(6 \beta_{0} \tilde{\eta}+\tilde{\eta}^2-8\right)\\
& \qquad \qquad \qquad \qquad +\beta_{0} \tilde{r} \tilde{\eta} (3 \beta_{0} \tilde{\eta}+2)-2 \tilde{r}^6-8 \tilde{r}^5 ]   \\
& R_{(0)(2)(0)(2)} = R_{(0)(3)(0)(3)} =\frac{1}{2 \tilde{r}^3 (\tilde{r}+1)^2 \tilde{\eta}^6} \left[ \beta_{0}^2 \tilde{\eta}^2+4 \tilde{r}^3 (\beta_{0} \tilde{\eta}+3)+\tilde{r}^2 \left(6 \beta_{0} \tilde{\eta}+\tilde{\eta}^2+4\right) \right. \nonumber \\
& \qquad \qquad \qquad \qquad \left.
+2 \beta_{0} \tilde{r} \tilde{\eta} (\beta_{0} \tilde{\eta}+1)+4 \tilde{r}^5+12 \tilde{r}^4 \right]
\\
& R_{(0)(2)(1)(2)} = R_{(0)(3)(1)(3)} =\frac{1}{\tilde{r} (\tilde{r}+1) \tilde{\eta}^5}\\
\nonumber
& R_{(2)(1)(1)(2)} = R_{(3)(1)(1)(3)} =\frac{1}{2 \tilde{r}^3 (\tilde{r}+1)^2 \tilde{\eta}^6} [\beta_{0}^2 \tilde{\eta}^2+\tilde{r}^2 \left(2 \beta_{0} \tilde{\eta}+\tilde{\eta}^2-8\right)+2 \beta_{0} \tilde{r} \tilde{\eta} (\beta_{0} \tilde{\eta}+1)
\\
& \qquad \qquad \qquad \qquad -8 \tilde{r}^5-24 \tilde{r}^4-24 \tilde{r}^3 ] \\
& R_{(2)(3)(2)(3)} =
\frac{1}{\tilde{r}^3 (\tilde{r}+1) \tilde{\eta}^6}
\left[\beta_{0}^2 \tilde{\eta}^2+\tilde{r}^2 (4-4 \beta_{0} \tilde{\eta})+\tilde{r} \tilde{\eta} (\tilde{\eta}-4 \beta_{0})+4 \tilde{r}^4+8 \tilde{r}^3 \right]
\end{align}
\end{subequations}

\section{Constraint on \texorpdfstring{$\lambda$}{\lambda} from energy momentum tensor}
\label{sec:lambda from EM tensor}

The energy-momentum tensor for a matter-field described by Lagrangian density $\mathcal{L}$ is:
\begin{align}
\label{eq:general-EM-tensor}
T_{\mu\nu} = -2 \left(\frac{\delta \mathcal{L}}{\delta g^{\mu\nu}}\right), \hspace{1cm} 
\end{align}
Substituting the action \eqref{eq:S_Balakin} for the non-minimally coupled electromagnetic field in the above expression, we have:
\begin{align}
\label{eq:NMC-EM-tensor}
\nonumber
T^{\mu}\, _{\nu} =  & F^{\mu\alpha} F_{\nu\alpha} - \frac{1}{4} F_{\alpha\beta} F^{\alpha\beta} g_{\eta\nu} g^{\mu\eta}  
+ \tilde{\lambda} \biggr[ -  \frac{3}{2} F^{\rho\alpha} g_{\nu\rho} F^{\beta\gamma} R_{\eta\alpha\beta\gamma} g^{\eta\mu}  - \frac{3}{2} F^{\mu\alpha} F^{\beta\gamma} R_{\nu\alpha\beta\gamma} 
\\
\nonumber
 & + \frac{1}{2} F^{\alpha\beta} F^{\gamma\rho} g_{\eta\nu} g^{\mu\eta} R_{\alpha\beta\gamma\rho} -   F_{\nu\rho}g^{\rho\alpha}\nabla_{\alpha}{\nabla_{\beta}{F^{\mu\beta}}}  
					   - F^{\mu\alpha} \nabla_{\alpha}{\nabla_{\beta}{F_{\nu\rho} g^{\rho\beta}}}  - 2 \nabla_{\alpha}{F^{\mu\alpha}}\nabla_{\beta}(F^{\eta\beta} g_{\eta\nu}) 
 \\
&			
- g_{\eta\nu} F^{\eta\alpha} \nabla_{\beta}(\nabla_{\alpha}{F^{\mu\beta}}) - F^{\mu\alpha} \nabla_{\beta}\left[\nabla_{\alpha}(F^{\eta\beta} g_{\eta\nu})\right] 	    - 2 \nabla_{\alpha}(F^{\eta\beta} g_{\eta\nu}) \nabla_{\beta}(F^{\mu\alpha}) \biggr] \, .
\end{align}  
The $00$ component of the stress-tensor is:
\begin{align}
\label{eq:NMC-EM00-tensor}
\nonumber
-\rho \equiv T^{0}\, _{0} =  & F^{0\alpha} F_{0\alpha} - \frac{1}{4} F_{\alpha\beta} F^{\alpha\beta} g_{\eta0} g^{0\eta}  
+ \tilde{\lambda} \biggr[ -  \frac{3}{2} F^{\rho\alpha} g_{0\rho} F^{\beta\gamma} R_{\eta\alpha\beta\gamma} g^{\eta0}  - \frac{3}{2} F^{0\alpha} F^{\beta\gamma} R_{0\alpha\beta\gamma} 
\\
\nonumber
 &+ \frac{1}{2} F^{\alpha\beta} F^{\gamma\rho} g_{\eta0} g^{0\eta} R_{\alpha\beta\gamma\rho} -   F_{0\rho}g^{\rho\alpha}\nabla_{\alpha}{\nabla_{\beta}{F^{0\beta}}}  
					   - F^{0\alpha} \nabla_{\alpha}{\nabla_{\beta}{F_{0\rho} g^{\rho\beta}}}  - 2 \nabla_{\alpha}{F^{0\alpha}}\nabla_{\beta}(F^{\eta\beta} g_{\eta0}) 
 \\
&			
 - g_{\eta0} F^{\eta\alpha} \nabla_{\beta}(\nabla_{\alpha}{F^{0\beta}}) - F^{0\alpha} \nabla_{\beta}\left[\nabla_{\alpha}(F^{\eta\beta} g_{\eta0})\right] 	    - 2 \nabla_{\alpha}(F^{\eta\beta} g_{\eta0}) \nabla_{\beta}(F^{0\alpha}) \biggr] \, ,
\end{align}
where $\rho$ is the energy-density and, by definition, it should be positive. 
For the zero electric field case for the Sultana-Dyer metric \eqref{eq:metric_SDyer_non-diagonal4}, the above $00$-th component becomes:
{
\begin{align}
\nonumber
T^{0}\,_{0} = & - (F_{\theta \tilde{r}}^{2} + F_{\phi \tilde{r}}^{2}) \left[\frac{ \tilde{r}-1}{2  \tilde{r}^3 \tilde{\eta}^8} +\tilde{\lambda}\left( \frac{- ( \tilde{r}-1) \tilde{\eta}^2+2  \tilde{r} \tilde{\eta}+4 \left(2  \tilde{r}^4+ \tilde{r}^2\right)}{ \tilde{r}^6 \tilde{\eta}^{14}}\right)\right] \\
%
%
& - F_{\phi \theta}^{2}\csc ^2(\theta )\, \left[\frac{1}{2  \tilde{r}^4 \tilde{\eta}^8} + \tilde{\lambda} \left( \frac{  \tilde{\eta}^2-4 \beta_{0}  \tilde{r} \tilde{\eta} +4 ( \tilde{r}+1)  \tilde{r}^2}{\tilde{r}^7 \tilde{\eta}^{14}}\right)\right]\label{eq:T00}
\end{align}
Demanding that $\rho > 0$ implies that the coefficients of $F_{\theta \tilde{r}}^2 + F_{\phi\tilde{r}}^2$ has to be $>0$. From coefficient of $F_{\theta \tilde{r}}^2 + F_{\phi\tilde{r}}^2$ we obtain,}
\begin{align}
\label{eq:condition-tilde-lambda-SD}
\tilde{\lambda}\, < \, \frac{ \tilde{r}^{3}\tilde{\eta}^6(\tilde{r}-1)}{-4\tilde{r}^2(1+2\tilde{r}^2) + \tilde{\eta}^{2}(\tilde{r} -1) -2\tilde{r}\tilde{\eta}}. 
\end{align}
Close to the horizon we can expand $\tilde{r} = ( 1 + {\color{black}\varepsilon}) $, where, ${\color{black}|\varepsilon|} < 1$. Subsituting this in the above inequality, we have
\begin{align}
\label{eq:constraint-lambda-SD}
\tilde{\lambda} < \tilde{r}^3\tilde{\eta}^4 \, \text{or,} \quad \tilde{\lambda} < \tilde{\eta}^4
\end{align}
Rewriting the above expression in terms of the cosmic time and setting 
$t = t_{0}$, or $\tilde{\eta} = \tilde{\eta}_0$ we get,
\begin{align}
\label{eq:constraint-lambda-2-rs}
\lambda \, < \, \rH^{2},
\end{align}
%
Similarly the coefficient of $F_{\phi\theta} > 0$ so that $\rho >0 $ and it leads to, 
\begin{align}
\tilde{\lambda} > -  \tilde{r}^3 \,\tilde{\eta}^4/2 
\end{align}
at $\tilde{\eta} = \tilde{\eta}_0$ we get, 
\begin{align}
\lambda > - \rH^2 
\end{align}
The final constraint we obtain, 
\begin{align}
- \frac{\rH^2}{2} <  \lambda < \rH^2 \label{eq:lambda-constraint-SD} 
\end{align}
%

\end{document}